\documentclass[%
 reprint,
superscriptaddress,
 amsmath,amssymb,
 aps, prl,
]{revtex4-2}

\usepackage{graphicx} 
\usepackage{dcolumn} 
\usepackage{bm} 
\usepackage{xcolor}
\usepackage[bookmarks=false,linkcolor=blue,urlcolor=blue,colorlinks,citecolor=blue]{hyperref}
\begin{document}

\preprint{APS/123-QED}

\title{Signatures of a topological phase transition in a planar Josephson junction}

\author{A.~Banerjee}
\affiliation{%
 Center for Quantum Devices, Niels Bohr Institute,
University of Copenhagen,
Universitetsparken 5, 2100 Copenhagen, Denmark
}

\author{O.~Lesser}
\affiliation{%
Department of Condensed Matter Physics, Weizmann Institute of Science, Rehovot, Israel 76100
}

\author{M.~A.~Rahman}
\affiliation{%
 Center for Quantum Devices, Niels Bohr Institute,
University of Copenhagen,
Universitetsparken 5, 2100 Copenhagen, Denmark
}

\author{H.-R.~Wang}
\affiliation{%
Department of Condensed Matter Physics, Weizmann Institute of Science, Rehovot, Israel 76100
}
\affiliation{%
 Department of Physics, Tsinghua University, Beijing 100084, China
}

\author{M.-R.~Li}
\affiliation{%
Department of Condensed Matter Physics, Weizmann Institute of Science, Rehovot, Israel 76100
}
\affiliation{%
 Department of Physics, Tsinghua University, Beijing 100084, China
}

\author{A.~Kringh{\o}j}
\affiliation{%
 Center for Quantum Devices, Niels Bohr Institute,
University of Copenhagen,
Universitetsparken 5, 2100 Copenhagen, Denmark
}

\author{A.~M.~Whiticar}
\affiliation{%
 Center for Quantum Devices, Niels Bohr Institute,
University of Copenhagen,
Universitetsparken 5, 2100 Copenhagen, Denmark
}

\author{A.~C.~C.~Drachmann}
\affiliation{%
 Center for Quantum Devices, Niels Bohr Institute,
University of Copenhagen,
Universitetsparken 5, 2100 Copenhagen, Denmark
}

\author{C. Thomas}
\affiliation{%
 Department of Physics and Astronomy,
Purdue University, West Lafayette, Indiana 47907 USA
}
\affiliation{ Birck Nanotechnology Center, Purdue University, West Lafayette, Indiana 47907 USA}

\author{T. Wang}
\affiliation{%
 Department of Physics and Astronomy,
Purdue University, West Lafayette, Indiana 47907 USA
}
\affiliation{ Birck Nanotechnology Center, Purdue University, West Lafayette, Indiana 47907 USA}

\author{M.~J.~Manfra}
\affiliation{%
Department of Physics and Astronomy,
Purdue University, West Lafayette, Indiana 47907 USA
}
\affiliation{ Birck Nanotechnology Center, Purdue University, West Lafayette, Indiana 47907 USA}
\affiliation{School of Materials Engineering, Purdue University, West Lafayette, Indiana 47907 USA}
\affiliation{School of Electrical and Computer Engineering, Purdue University, West Lafayette, Indiana 47907 USA}

\author{E.~Berg}
\affiliation{%
Department of Condensed Matter Physics, Weizmann Institute of Science, Rehovot, Israel 76100
}

\author{Y.~Oreg}
\affiliation{%
Department of Condensed Matter Physics, Weizmann Institute of Science, Rehovot, Israel 76100
}

\author{Ady~Stern}
\affiliation{%
Department of Condensed Matter Physics, Weizmann Institute of Science, Rehovot, Israel 76100
}

\author{C.~M.~Marcus}
\affiliation{%
 Center for Quantum Devices, Niels Bohr Institute,
University of Copenhagen,
Universitetsparken 5, 2100 Copenhagen, Denmark
}


\begin{abstract}
A growing body of work suggests that planar Josephson junctions fabricated using superconducting hybrid materials provide a highly controllable route toward one-dimensional topological superconductivity.  Among the experimental controls are in-plane magnetic field, phase difference across the junction, and carrier density set by electrostatic gate voltages. Here, we investigate planar Josephson junctions with an improved design based on an epitaxial InAs/Al heterostructure, embedded in a superconducting loop, probed with integrated quantum point contacts (QPCs) at both ends of the junction. For particular ranges of in-plane field and gate voltages, a closing and reopening of the superconducting gap is observed, along with a zero-bias conductance peak (ZBCP) that appears upon reopening of the gap. Consistency with a simple theoretical model supports the interpretation of a topological phase transition. While gap closings and reopenings generally occurred together at the two ends of the junction, the height, shape, and even presence of ZBCPs typically differed between the ends, presumably due to disorder and variation of couplings to local probes.
\end{abstract}

\maketitle

\begin{figure}[t!]
\includegraphics[width=0.5\textwidth]{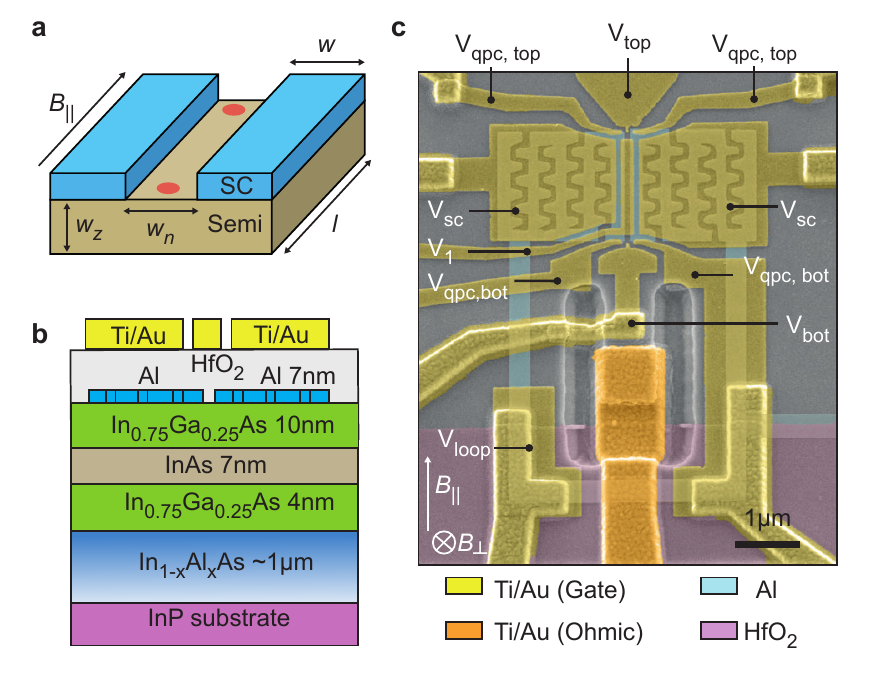}
\caption{\label{fig01}{{\bf Planar Josephson junction device.} (a)~Schematic of a planar Josephson junction consisting of two superconducting leads (blue) in epitaxial contact with the underlying semiconductor (brown). Between the leads of width $w = 1.8 \,\mu$m was a semiconductor (normal) region of width $w_n = 100$~nm and length $l= 1.6 \,\mu$m. The nominal thickness $w_z\sim 20$~nm of the active semiconductor region contains two barriers and the InAs quantum well. Red dots schematically indicate the positions of theoretically predicted Majorana zero modes in the topological phase. (b)~Schematic cross section shows the Al/InAs heterostructure with layer thicknesses along with dielectric and gate layers. Leads and junction were covered by 15~nm of HfO$_2$ dielectric deposited by atomic layer deposition and Ti/Au electrostatic gates. (c)~False-color electron micrograph of a representative device. The superconducting leads have meandering perforations to allow partial depletion below using gate voltage $V_{\rm SC}$. Leads are connected by a superconducting loop allowing phase biasing of the junction using a small out-of-plane magnetic field $B_{\perp}$. Tunnelling spectroscopy is performed using quantum point contacts at the junction ends, controlled by voltages $V_{\rm qpc,top}$ and $V_{\rm top}$ on the top and $V_{\rm qpc,bot}$ and $V_{\rm bot}$ on the bottom.}}
\end{figure}

Planar superconductor-normal-superconductor (SNS) Josephson junctions (JJs) with sufficient spin-orbit coupling can exhibit one-dimensional topological superconductivity in the presence of a magnetic field applied parallel to the SN interfaces. Theoretically, the N region under these conditions acts as a quasi-one-dimensional topological wire bounded by trivial superconducting walls, with Majorana zero modes at its ends~\cite{HellFlensberg,Pientka,Setiawan}. Compared to alternative nanowire platforms~\cite{Oreg,Lutchyn,Mourik,Das,Deng}, planar JJs have a new experimental knob, the phase difference between bounding trivial superconductors, which can lower the magnetic field required to observe a topological phase transition, as reported in recent experiments in Al/InAs \cite{fornieri,ShabaniReopening}, Al/HgTe~\cite{ren} and NbTiN/InSb~\cite{GoswamiReopening}. 

Previous studies on related structures \cite{fornieri} demonstrated the formation of a zero-bias conductance peak (ZBCP) at one end of an Al/InAs planar JJ device. The parallel magnetic field, $B_\parallel$, at which the ZBCP first appeared depended on the phase difference, $\phi$, across the junction, first appearing at $\phi \sim \pi$, as expected for a topological phase transition~\cite{HellFlensberg,Pientka}. A related effect was reported by Ren~{\it et al.}~\cite{ren}, who found that the
ZBCP appears in a diamond-shaped region in the $\phi$--$B_\parallel$ plane. 
Ke~{\it et al.} observed an expected minimum of critical current at a gate-voltage dependent value of $B_\parallel$~\cite{GoswamiReopening}. Dartiailh~{\it et al.} reported a similar signature and additionally detected a $\pi$ phase shift of the current-phase relation associated with revival of the supercurrent~\cite{ShabaniReopening}. 

Here, we extend our previous investigation of topological superconductivity in planar JJs \cite{fornieri}  using an improved design that helps preserve the hard superconducting gap in the leads in the presence of $B_\parallel$, allowing wide leads~\cite{Setiawan}. The junction is embedded in a  superconducting loop, allowing controlled biasing of $\phi$ using externally applied flux, and the junction region can now be probed at 
both ends via tunnelling spectroscopy using quantum point contacts (QPCs). In tuned ranges of junction gate voltage, we observe a closing and reopening of the superconducting gap with increasing $B_\parallel$, along with a concurrent appearance of a ZBCP at one or both ends of the junction. The gap reopening and the appearance of a ZBCP both depend on $\phi$ and remain concurrent when $\phi$ is modulated by flux. 

To test our interpretation of these observations in terms of a topological phase transition, we investigate a simple model of the system that includes spin-orbit coupling as well as both Zeeman and orbital effects of the in-plane magnetic field. The orbital effect is due to the finite thickness of the Al-InAs heterostructure stack. As discussed below (see Supplementary Material: Methods, Fig.~S\ref{suppFigS2}a, and Fig.~S\ref{suppFigS11}), for realistic parameters, the model exhibits a topological phase for $\sim 10\%$ of parameter space examined, showing many features observed in the experiment. The model also shows non-topological near-closings of the gap. In the experiment, a similar fraction, around $10\%$, of junction gate voltages showed a ZBCP following the gap reopening.  

Planar JJ devices were fabricated using an InAs-based heterostructure grown on an InP wafer, with epitaxial Al as the topmost layer of the heterostructure [see Fig.~\ref{fig01}(b)]. In$_{0.75}$Ga$_{0.25}$As barriers separate the InAs quantum well from the Al layer above and the In$_{1-x}$Al$_{x}$As graded buffer below. The JJ and  superconducting loop were fabricated by a combination of selective wet etching of Al (using Transene D etchant) and deep wet etching of the heterostructure stack to form a mesa and U-shaped trench. A Ti/Au layer contacting a patch of the mesa (with Al removed) serves as a sub-micron internal ohmic contact allowing bottom-end tunnelling spectroscopy through a QPC inside the superconducting loop. Patterned HfO$_2$  dielectric was deposited using atomic layer deposition (ALD) to allow the Ti/Au layer contacting the internal ohmic contact to pass over the superconducting loop. A second layer of ALD HfO$_2$ was then deposited on the entire chip followed by patterned deposition of Ti/Au gates to electrostatically control the junction and QPCs. The JJ (width $w_n=100$~nm, length $l=1.6\,\mu$m) was covered by a gate above the second ALD layer, energized by  gate voltage $V_1$ relative to the leads to control carrier density and mean free path in the junction [Fig.~\ref{fig01}(c)]. Dependence of density and mobility on gate voltage was investigated in a Hall-bar geometry made from the same material, with similar dielectric and top gate (see Supplementary Material Fig.~S\ref{suppFigS12}).

The Al layer in the leads (width $w=1.8$~$\mu$m) was etched to form meandering perforations (width $\sim100$~nm). These perforations allowed depletion of the semiconductor below and laterally when the gate voltage covering the leads was set to a large negative value, $V_{\rm SC} \sim -3$~V. Depletion in the meanders resulted in an improved hard superconducting gap up to $B_\parallel \sim 0.5$~T (see Supplementary Material Fig.~S\ref{suppFigS1} for tunnelling spectroscopy in a lead-like structure). The two leads are connected through a superconducting loop (with undepleted electron gas below) with area $\sim$12~$\mu$m$^2$ allowing phase biasing of the junction by the application of a perpendicular magnetic field, $B_{\perp}$. One flux quantum, $\Phi_0=h/2e$, through the loop corresponds to  $B_\perp \sim 0.17$~mT, small compared to the field that closes the induced gap under the Al ($B_{\perp} \sim 10\,$mT) or that drives the Al normal ($B_{\perp} \sim 40\,$mT). Split gates controlled by voltages $V_{\rm qpc,top}$ and $V_{\rm qpc,bot}$ electrostatically define constrictions at the top and bottom of the junction to serve as QPC tunnel barriers. Gate voltages $V_{\rm top}$ and $V_{\rm bot}$, which control densities in the normal regions outside the QPCs, are typically fixed at $\sim +100$~mV.  We show results from four devices of identical design.  We first focus on tunnelling spectra from the top end of the junction for Devices 1 and 2, and then examine spectra measured simultaneously at both ends of the junction for Devices 3 and 4.  

\begin{figure}
\includegraphics[width=0.5\textwidth]{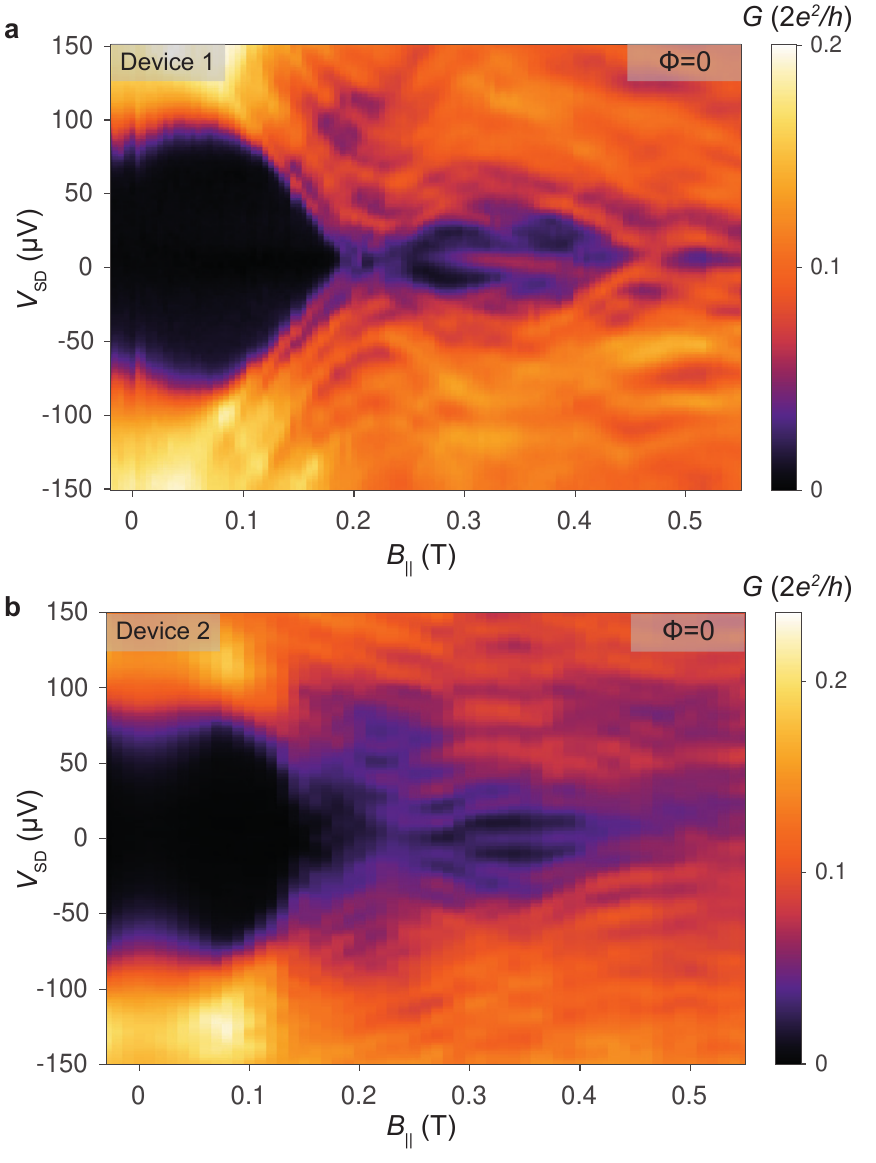} 
\caption{\label{fig02} {\bf Tunnelling spectroscopy as a function of in-plane magnetic field.} Differential conductance, $G$, as a function of source-drain bias $V_{\rm SD}$ and magnetic field $B_{\parallel}$ along the junction, showing a closing of the superconducting gap followed by reopening and concurrent appearance of a ZBCP in (a)~Device 1, with $V_1=+86$~mV, $V_{\rm SC}=-1.5$~V, $V_{\rm qpc,top}=-0.31$~V, $V_{\rm qpc,bot}=-3.0$~V, $V_{\rm loop}=-3.0$~V.  (b)~Device 2, with $V_1=+185$~mV, $V_{\rm SC}=-6.0$~V, $V_{\rm qpc,top}=-0.48$~V, $V_{\rm qpc,bot}=-1.61$~V, $V_{\rm loop}=-3.0$~V. }
\end{figure}

Figure~\ref{fig02} shows differential conductance, $G$, as a function of source-drain bias $V_{\rm SD}$ measured at the top of the junction (outside the loop) as a function of $B_{\parallel}$ applied along the junction for two devices. To compensate spurious flux through the superconducting loop due to sample misalignment, $G$ was measured as a function of $B_\perp$ at each value of $B_{\parallel}$ and reconstructed to plot the $B_{\parallel}$ dependence at fixed flux (see Methods). Figure~\ref{fig02} is for the case of zero flux, $\Phi=0$. Top QPC gates were tuned to operate in the tunnelling regime, where $G$ is roughly proportional to the local density of states (see \cite{tinkham_introduction_2004}, Sec.~11.5).  

At $B_{\parallel}=0$, we measured a gap $\Delta \sim 80~\mu$eV, which increased to $\Delta \sim 100~\mu$eV at $B_\parallel \sim 0.05$~T. Above 0.1~T, a dense but striated set of tunnelling peaks approach zero bias, closing at $B_\parallel \sim 0.2$~T. With further increase of field, the gap reopened, and a ZBCP appeared, separated from the gapped states. A maximum gap of $\sim 20-30~\mu$eV was observed in the reopened state before it closed again at $B_\parallel \sim 0.5$~T.

\begin{figure}
\includegraphics[width=0.5\textwidth]{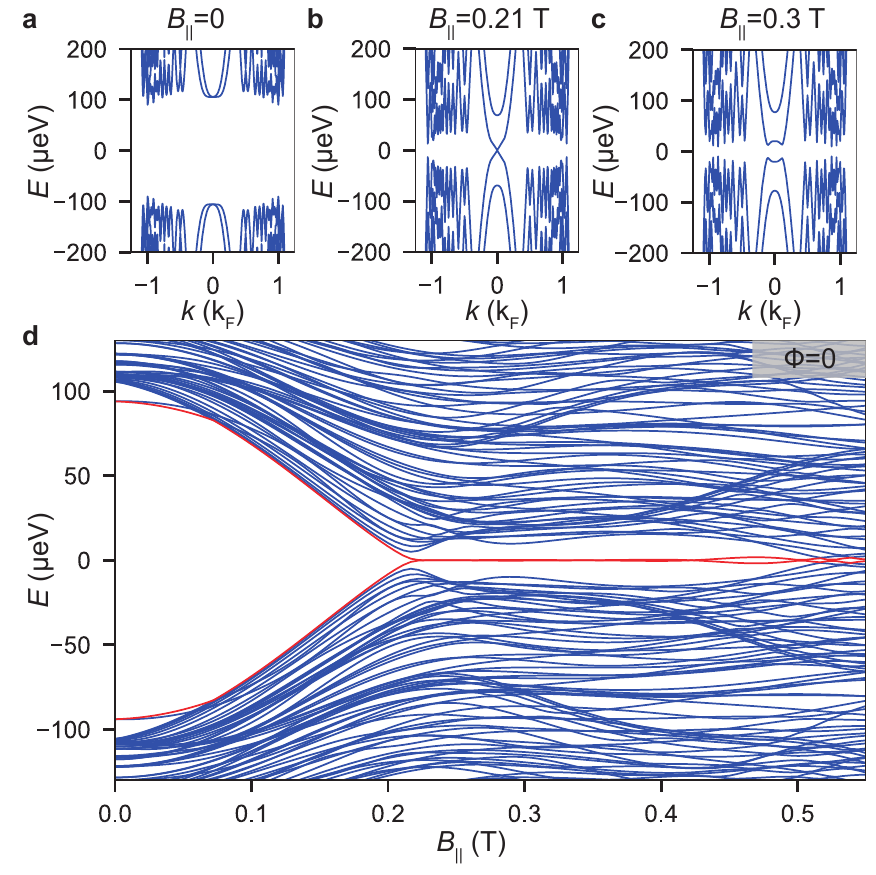} 
\caption{\label{fig03}  {\bf Theoretical model of topological phase transition.} Dispersion  of the Andreev bound states in a Josephson junction with periodic boundary conditions as a function of momentum $k$ along the junction (measured in units of $k_{\rm F}=\sqrt{2m^{*}\mu_{\rm N}}/\hbar$) at phase difference $\phi=0$ for three different values of the Zeeman field: (a)~The spectrum is fully gapped at $B_\parallel=0$. (b)~At $B_\parallel= $~0.21 T, the gap at $k=0$ closes. (c)~At $B_\parallel = 0.3$~T, the gap at $k=0$ has reopened, implying a topologically inverted superconducting gap. The gap at non-zero momentum remains non-zero throughout. (d)~Andreev bound state spectrum of a finite-length planar Josephson junction ($l=4\,\mu$m) with open boundary conditions. The closing and reopening of the superconducting gap at $B_\parallel = $~0.21 T is followed by the appearance of a Majorana state at zero energy (red), signaling a transition to the topological phase.}
\end{figure}

\begin{figure*}[t!]
\includegraphics[width=1\textwidth]{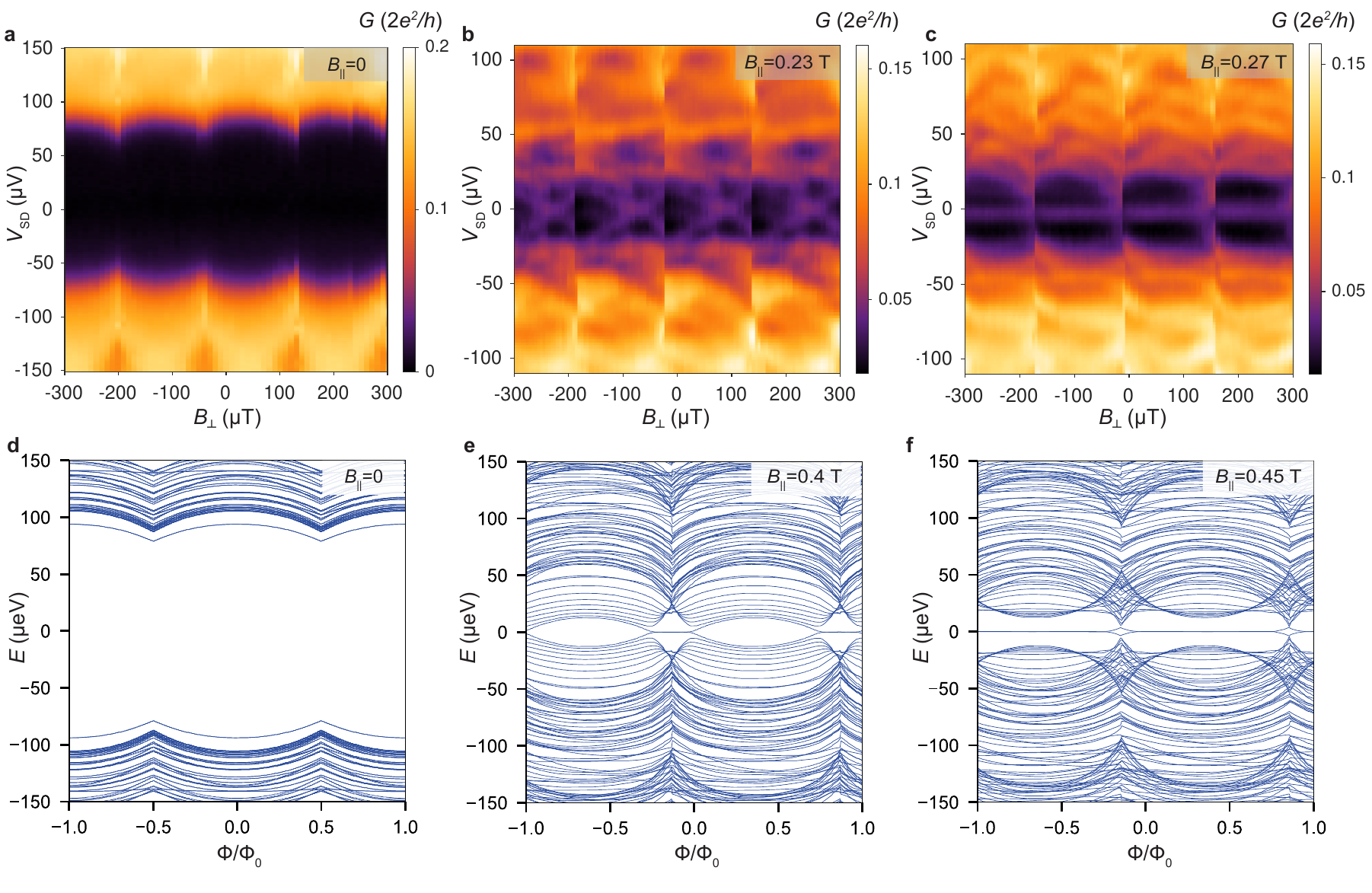} 
\caption{\label{fig04} {\bf Flux dependence.}  Differential conductance, $G$, as a function of source-drain bias $V_{\rm SD}$ and out-of-plane magnetic field $B_{\perp}$ penetrating the flux loop, at different values of in-plane magnetic field $B_{\parallel}$. (a) At $B_{\parallel}=$~0, the superconducting gap is modulated periodically as a function of $B_{\perp}$. The period corresponds to $\Phi_0 = h/2e$ through the superconducting loop. (b) At $B_{\parallel}=$~0.2~T, states cross zero energy in a bowtie shape, indicating a phase-dependent gap closing. (c)~At $B_{\parallel}=$~0.27~T the superconducting gap reopens with a stable ZBCP. (d)--(f) Theoretical spectra as a function of the flux at three values of $B_{\parallel}=$~0, 0.4~T and 0.45~T (note, these are not the same fields as the experimental data in a-c, suggesting only qualitative correspondence). The simulations take into account the inductance $L=$~2 nH of the flux loop.}
\end{figure*}

We compare these experimental observations to a theoretical model, extending models developed in Refs.~\cite{Pientka,HellFlensberg}. The proximity-coupled semiconductor is treated as a parabolic band, approximated within a tight-binding model, with effective mass of $m^* = 0.026\,m_e$, where $m_e$ is the free electron mass, and Rashba spin-orbit coupling $\alpha = 15$~meV~nm.  The superconducting leads are represented by a pairing potential $\Delta \sim 140$~$\mu$eV. The in-plane field $B_\parallel$ induces both a Zeeman coupling and an orbital effect. The Zeeman coupling is characterized by an energy scale $E_{\rm Z}=  g_{{\rm S(N)}} \mu_{\rm B} B_\parallel$/2, where $\mu_{\rm B}$ is the Bohr magneton, with g-factors $g_{\rm N}=-8$ in the junction and $g_{\rm S}=-4$ in the leads, based on literature values~\cite{lee2019transport,NicheleScaling}. The orbital effect is due to the finite cross section of the device, $w_z(w_n + 2\xi)\sim(20\,{\rm nm})(0.5\,\mu{\rm m})\sim \Phi_0/0.2$~T, where $\xi\sim$~200~nm is the superconducting coherence length. As discussed below, this orbital field scale emerges naturally in the model and is not put in by hand. The orbital effect is included by considering a bilayer structure with complex hopping between layers~\cite{peierls_zur_1933} and a linearly increasing superconducting phase difference between the layers and across the junction (see~\cite{tinkham_introduction_2004}, Sec.~6.4 and \cite{bennemann2008superconductivity}, Sec.~2.9).

In the model, the quasi-one-dimensional junction supports Andreev bound states with  momentum dispersion as shown in Fig.~\ref{fig03}, where $k$ is momentum parallel to the SN interfaces. At zero field [Fig.~\ref{fig03}(a)], the spectrum shows a momentum-dependent superconducting gap that is induced by lateral proximity effect from the leads. At $B_{\parallel} \sim 0.2$~T a topological phase transition occurs, signaled by a closing of the gap at $k=0$ [Fig.~\ref{fig03}(b)]. Increasing $B_\parallel$ further reopens the gap, as illustrated in Fig.~\ref{fig03}(c) for the case $B_{\parallel}=0.3$~T. Notice that the spectrum remains gapped at finite $k \sim \pm k_{\rm F}$ throughout this field range. Correspondingly, in a Josephson junction with open boundary conditions, the bulk remains gapped away from the transition point. Figure~\ref{fig03}(d) shows the model spectrum in a finite-length junction undergoing a gap closing at $B_\parallel \sim $~0.2~T and reopening, accompanied by the appearance of a zero-energy state. The zero-energy state observed in the model corresponds to a Majorana zero mode. While the gap closure around $B_\parallel \sim $~0.2~T is robust, i.e., insensitive to small changes in chemical potential, this feature  can  be associated either with a topological transition accompanied by zero-energy states or with a near-closing without a topological transition, depending on relatively small changes in chemical potential or other model parameters. This is shown in Supplementary Material Fig.~S\ref{suppFigS2}.

 \begin{figure*}[t!]
\includegraphics[width=1\textwidth]{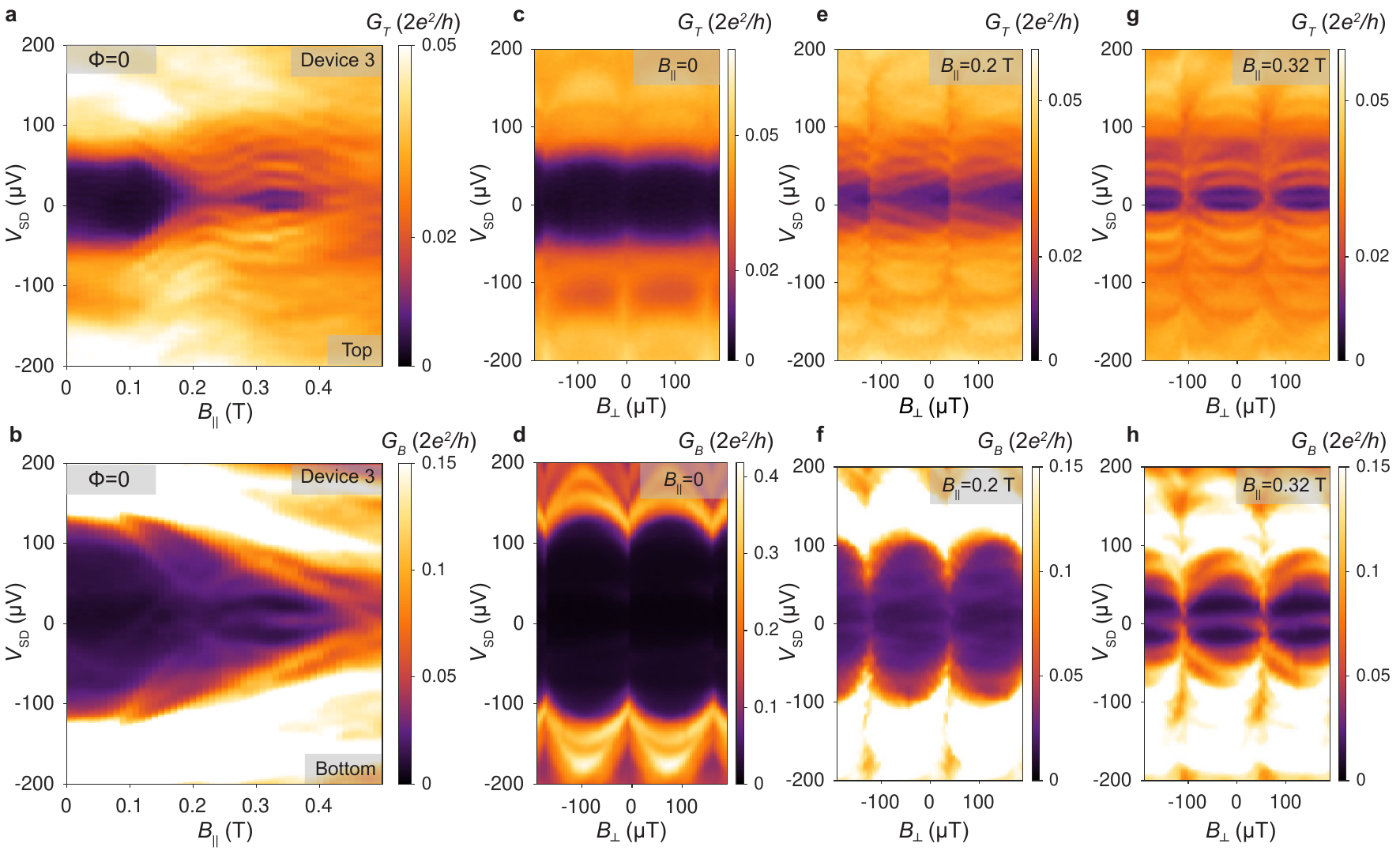} 
 \caption{\label{fig05} {\bf Two-ended tunnelling spectroscopy at the two ends of the junction. } Differential conductance measured as a function of source-drain bias $V_{\rm SD}$ and in-plane magnetic field $B_{\parallel}$ (a)~$G_{T}$~the top end and (b)~$G_{B}$~at the bottom end. The phase bias is set to $\Phi=0$. Both ends display a closing and reopening of the gap at $B_{\parallel} \sim $0.22~T followed by a zero-bias conductance peak. Simultaneous differential conductance measured at the top end and bottom end as a function of source-drain bias $V_{\rm SD}$ and out-of-plane magnetic field $B_{\perp}$ for different values of in-plane magnetic field $B_{\parallel}$. (c)~and (d)~At $B_{\parallel}=$~0, the superconducting gap is modulated periodically at both ends as a function of $B_{\perp}$ with the same periodicity and zero relative phase difference. (e) and (f) At $B_{\parallel}=$~0.2~T,  the spectrum at both ends becomes gapless for all values of $B_{\perp}$.  (g) and (h) At $B_{\parallel}=$~0.3~T the superconducting gap reopens with a stable zero-bias conductance peak at both ends of the device. Gate voltages were $V_1=+189$~mV, $V_{\rm SC}=-2.6$~V, $V_{\rm qpc,top}=-6$~mV,$V_{\rm top}=-0.1$~V, $V_{\rm qpc,bot}=-265$~mV, $V_{\rm bot}=+0.2$~V, and $V_{\rm loop}=-3.0$~V.}
\end{figure*}

We next examine the effects of phase bias on subgap spectroscopy. Figures~\ref{fig04}(a-c) show tunnelling spectra as a function of  $B_\perp$ at different values of $B_\parallel$ in Device 1.
At  $B_\parallel = 0$ [Fig.~\ref{fig04}(a)], the induced superconducting gap is modulated periodically as a function of $B_\perp$ with a periodicity of $\Delta B_\perp \sim 170~\mu$T, corresponding to one  flux quantum $\Phi_0=h/2e$  through the loop. The maximum (minimum) induced  gap is  $\Delta \sim 80~\mu$eV ($50\mu$eV) at integer (half-integer) flux through the loop. Around half flux quantum values, sharp switches are observed, which we attribute to phase slips due to the large inductance of the loop, $L \sim 2$~nH \cite{NicheleRelating} (see Methods). 

Increasing $B_\parallel$ from zero, the phase-dependent states initially moved to lower energy up to the first gap closing. Figure~\ref{fig04}(b) shows the phase-dependent spectrum at $B_\parallel = 0.23$~T, corresponding to the first gap closing in Fig.~\ref{fig02}(a). Within each flux lobe, a bowtie-shaped set of states crossing zero energy was observed, creating a gapless spectrum. When the in-plane field was increased to $B_\parallel=0.27$~T, the gap reappeared along with a ZBCP [Fig.~\ref{fig04}(c)]. The ZBCP displays no observable dependence on $B_\perp$, while the gap shows strong phase dependence with the lowest-lying energy level at $E \sim 30~\mu$eV. In contrast to the phase-dependent spectrum at $B_\parallel =0$ [Fig.~\ref{fig04}(a)], spectra at finite parallel field [Figs.~\ref{fig04}(b-c)] are asymmetric in phase bias within each lobe~\cite{Avignon,Nazarov,Pientka, Tosi}. 

 The numerical bound-state spectrum was determined as a function of $\Phi$, including the effect of loop inductance (see Methods). Figures~\ref{fig04}(d-f), show  numerical spectra with variation of the magnetic flux at three values of $B_\parallel$.
 At zero in-plane magnetic field [Fig.~\ref{fig04}(d)], the spectrum is spin degenerate and all Andreev bound state energies are periodically modulated as a function of $\Phi$. 
 At intermediate magnetic fields [Fig.~\ref{fig04}(e)], the system is trivial in some range of $\Phi$ and topological in another range. In the topological region, a zero-energy state appears in the gap. 
 These regions are separated by a gap-closing transition. For higher magnetic fields [Fig.~\ref{fig04}(f)], the spectrum becomes topological for all values of $\Phi$ and the junction hosts a stable zero-energy state.
 
 We next examine the effect on the spectrum when  $B_\parallel$ was tilted by small angles in the plane of the junction. As shown in Figs.~S\ref{suppFigS7}(a-c), a tilt angle of $\sim20^\circ$ closed the reopened gap. Similar behavior is seen in the model, though with greater sensitivity to tilt, as seen in Figs.~S\ref{suppFigS7}(d-f). 

Finally, we investigate simultaneous tunnelling spectroscopy at both ends of the device using the two QPCs in Device 3. Differential conductances $G_{\rm T}$ and $G_{\rm B}$, measured at the top (top row of Fig.~\ref{fig05}) and bottom (bottom row of Fig.~\ref{fig05}) of the junction show correlated modulation of the superconducting gaps at the two ends as a function of $B_\perp$ with flux switches occurring at the same values of $B_\perp$ at both ends [Figs.~\ref{fig05}(c) and ~\ref{fig05}(d)]. The sizes of the superconducting gaps at the two ends are different, with $\Delta_{\rm T} \sim$ 50~$\mu$eV at the top and $\Delta_{\rm B} \sim$~120 $\mu$eV at the bottom. In the presence of an in-plane magnetic field, the gaps at the two ends disappeared simultaneously at $B_\parallel \sim 0.2~$T~before reopening and undergoing a final gap closure at $B_\parallel \sim 0.48$~T. A gapless spectrum at  $B_\parallel \sim 0.2$~T was seen for all values of $B_\perp$ [Figs.~\ref{fig05}(e,f)]. At $B_\parallel \sim 0.3$~T, the maximal reopened gaps at the two ends have different magnitudes, with a smaller gap at the top end ($\Delta_{\rm T} \sim$ 30~$\mu$eV) compared to the bottom end ($\Delta_{\rm B} \sim$ 50~$\mu$eV). Both ends display ZBCPs that emerge from the gap-reopening and were reasonably stable for a range of in-plane magnetic field, phase [Figs.~\ref{fig05}(g,h)] as well as junction gate voltage gate $V_1$. However, their range in $V_1$ was not strongly correlated (see Supplementary Material Fig.~S\ref{suppFigS8}).

The observed ZBCPs emerging from a gap reopening are consistent with the model, which exhibits topological superconductivity. However, the lack of general end-to-end correlations suggest the importance of long-wavelength disorder as well as sensitivity of coupling of subgap states to local probes. We speculate that the strength of disorder in our devices is low enough to allow observation of a simultaneous reopening of the gap on the ends, as the closing and reopening is a bulk property that does not depend sensitively on local probe coupling. On the other hand, ZBCPs in local measurements are sensitive to the details of couplings, and so ought to be less robust with varying parameters and disorder. Experiments that simultaneously detect both local and bulk properties, such as non-local conductance, will help resolve this matter \cite{Akhmerov,DanonNonlocal,MenardNonlocal,PugliaNonlocal,Sarma3T}.

\bibliography{Gap_reopening.bib}

\begin{thebibliography}{10}
\expandafter\ifx\csname url\endcsname\relax
  \def\url#1{\texttt{#1}}\fi
\expandafter\ifx\csname urlprefix\endcsname\relax\def\urlprefix{URL }\fi
\providecommand{\bibinfo}[2]{#2}
\providecommand{\eprint}[2][]{\url{#2}}

\bibitem{HellFlensberg}
\bibinfo{author}{Hell, M.}, \bibinfo{author}{Leijnse, M.} \&
  \bibinfo{author}{Flensberg, K.}
\newblock \bibinfo{title}{Two-dimensional platform for networks of {Majorana}
  bound states}.
\newblock \emph{\bibinfo{journal}{Phys. Rev. Lett.}}
  \textbf{\bibinfo{volume}{118}}, \bibinfo{pages}{107701}
  (\bibinfo{year}{2017}).

\bibitem{Pientka}
\bibinfo{author}{Pientka, F.} \emph{et~al.}
\newblock \bibinfo{title}{Topological superconductivity in a planar {Josephson}
  junction}.
\newblock \emph{\bibinfo{journal}{Phys. Rev. X}} \textbf{\bibinfo{volume}{7}},
  \bibinfo{pages}{021032} (\bibinfo{year}{2017}).

\bibitem{Setiawan}
\bibinfo{author}{Setiawan, F.}, \bibinfo{author}{Stern, A.} \&
  \bibinfo{author}{Berg, E.}
\newblock \bibinfo{title}{Topological superconductivity in planar {Josephson}
  junctions: Narrowing down to the nanowire limit}.
\newblock \emph{\bibinfo{journal}{Phys. Rev. B}} \textbf{\bibinfo{volume}{99}},
  \bibinfo{pages}{220506} (\bibinfo{year}{2019}).

\bibitem{Oreg}
\bibinfo{author}{Oreg, Y.}, \bibinfo{author}{Refael, G.} \&
  \bibinfo{author}{von Oppen, F.}
\newblock \bibinfo{title}{Helical liquids and {{Majorana}} bound states in
  quantum wires}.
\newblock \emph{\bibinfo{journal}{Phys. Rev. Lett.}}
  \textbf{\bibinfo{volume}{105}}, \bibinfo{pages}{177002}
  (\bibinfo{year}{2010}).

\bibitem{Lutchyn}
\bibinfo{author}{Lutchyn, R.~M.}, \bibinfo{author}{Sau, J.~D.} \&
  \bibinfo{author}{Das~Sarma, S.}
\newblock \bibinfo{title}{{Majorana} fermions and a topological phase
  transition in semiconductor-superconductor heterostructures}.
\newblock \emph{\bibinfo{journal}{Phys. Rev. Lett.}}
  \textbf{\bibinfo{volume}{105}}, \bibinfo{pages}{077001}
  (\bibinfo{year}{2010}).

\bibitem{Mourik}
\bibinfo{author}{Mourik, V.} \emph{et~al.}
\newblock \bibinfo{title}{Signatures of {{Majorana}} fermions in hybrid
  superconductor-semiconductor nanowire devices}.
\newblock \emph{\bibinfo{journal}{Science}} \textbf{\bibinfo{volume}{336}},
  \bibinfo{pages}{1003--1007} (\bibinfo{year}{2012}).

\bibitem{Das}
\bibinfo{author}{Das, A.} \emph{et~al.}
\newblock \bibinfo{title}{Zero-bias peaks and splitting in an {Al}--{InAs}
  nanowire topological superconductor as a signature of {{Majorana}} fermions}.
\newblock \emph{\bibinfo{journal}{Nature Physics}}
  \textbf{\bibinfo{volume}{8}}, \bibinfo{pages}{887--895}
  (\bibinfo{year}{2012}).

\bibitem{Deng}
\bibinfo{author}{Deng, M.} \emph{et~al.}
\newblock \bibinfo{title}{{Majorana} bound state in a coupled quantum-dot
  hybrid-nanowire system}.
\newblock \emph{\bibinfo{journal}{Science}} \textbf{\bibinfo{volume}{354}},
  \bibinfo{pages}{1557--1562} (\bibinfo{year}{2016}).

\bibitem{fornieri}
\bibinfo{author}{Fornieri, A.} \emph{et~al.}
\newblock \bibinfo{title}{Evidence of topological superconductivity in planar
  {Josephson} junctions}.
\newblock \emph{\bibinfo{journal}{Nature}} \textbf{\bibinfo{volume}{569}},
  \bibinfo{pages}{89--92} (\bibinfo{year}{2019}).

\bibitem{ShabaniReopening}
\bibinfo{author}{Dartiailh, M.~C.} \emph{et~al.}
\newblock \bibinfo{title}{Phase signature of topological transition in
  {Josephson} junctions}.
\newblock \emph{\bibinfo{journal}{Phys. Rev. Lett.}}
  \textbf{\bibinfo{volume}{126}}, \bibinfo{pages}{036802}
  (\bibinfo{year}{2021}).

\bibitem{ren}
\bibinfo{author}{Ren, H.} \emph{et~al.}
\newblock \bibinfo{title}{Topological superconductivity in a phase-controlled
  {Josephson} junction}.
\newblock \emph{\bibinfo{journal}{Nature}} \textbf{\bibinfo{volume}{569}},
  \bibinfo{pages}{93--98} (\bibinfo{year}{2019}).

\bibitem{GoswamiReopening}
\bibinfo{author}{Ke, C.~T.} \emph{et~al.}
\newblock \bibinfo{title}{Ballistic superconductivity and tunable
  $\pi$--junctions in {InSb} quantum wells}.
\newblock \emph{\bibinfo{journal}{Nature communications}}
  \textbf{\bibinfo{volume}{10}}, \bibinfo{pages}{1--6} (\bibinfo{year}{2019}).

\bibitem{tinkham_introduction_2004}
\bibinfo{author}{Tinkham, M.}
\newblock \emph{\bibinfo{title}{Introduction to Superconductivity}}.
\newblock International Series in Pure and Applied Physics
  (\bibinfo{publisher}{{McGraw-Hill}}, \bibinfo{address}{{New York}},
  \bibinfo{year}{1996}), \bibinfo{edition}{2nd} edn.

\bibitem{lee2019transport}
\bibinfo{author}{Lee, J.} \emph{et~al.}
\newblock \bibinfo{title}{Transport studies of epi-{A}l/{I}n{A}s
  two-dimensional electron gas systems for required building-blocks in
  topological superconductor networks}.
\newblock \emph{\bibinfo{journal}{Nano Letters}} \textbf{\bibinfo{volume}{19}},
  \bibinfo{pages}{3083} (\bibinfo{year}{2019}).

\bibitem{NicheleScaling}
\bibinfo{author}{Nichele, F.} \emph{et~al.}
\newblock \bibinfo{title}{Scaling of {{Majorana}} zero-bias conductance peaks}.
\newblock \emph{\bibinfo{journal}{Phys. Rev. Lett.}}
  \textbf{\bibinfo{volume}{119}}, \bibinfo{pages}{136803}
  (\bibinfo{year}{2017}).

\bibitem{peierls_zur_1933}
\bibinfo{author}{Peierls, R.}
\newblock \bibinfo{title}{{Zur Theorie des Diamagnetismus von
  Leitungselektronen}}.
\newblock \emph{\bibinfo{journal}{Zeitschrift f\"ur Physik}}
  \textbf{\bibinfo{volume}{80}}, \bibinfo{pages}{763--791}
  (\bibinfo{year}{1933}).

\bibitem{bennemann2008superconductivity}
\bibinfo{author}{Bennemann, K.-H.} \& \bibinfo{author}{Ketterson, J.~B.}
\newblock \emph{\bibinfo{title}{Superconductivity: Volume 1: Conventional and
  Unconventional Superconductors Volume 2: Novel Superconductors}}
  (\bibinfo{publisher}{Springer Science \& Business Media},
  \bibinfo{year}{2008}).

\bibitem{NicheleRelating}
\bibinfo{author}{Nichele, F.} \emph{et~al.}
\newblock \bibinfo{title}{Relating {Andreev} bound states and supercurrents in
  hybrid {Josephson} junctions}.
\newblock \emph{\bibinfo{journal}{Phys. Rev. Lett.}}
  \textbf{\bibinfo{volume}{124}}, \bibinfo{pages}{226801}
  (\bibinfo{year}{2020}).

\bibitem{Avignon}
\bibinfo{author}{Reynoso, A.~A.}, \bibinfo{author}{Usaj, G.},
  \bibinfo{author}{Balseiro, C.~A.}, \bibinfo{author}{Feinberg, D.} \&
  \bibinfo{author}{Avignon, M.}
\newblock \bibinfo{title}{Spin-orbit-induced chirality of {Andreev} states in
  {Josephson} junctions}.
\newblock \emph{\bibinfo{journal}{Phys. Rev. B}} \textbf{\bibinfo{volume}{86}},
  \bibinfo{pages}{214519} (\bibinfo{year}{2012}).

\bibitem{Nazarov}
\bibinfo{author}{Yokoyama, T.}, \bibinfo{author}{Eto, M.} \&
  \bibinfo{author}{Nazarov, Y.~V.}
\newblock \bibinfo{title}{Anomalous {Josephson} effect induced by spin-orbit
  interaction and zeeman effect in semiconductor nanowires}.
\newblock \emph{\bibinfo{journal}{Phys. Rev. B}} \textbf{\bibinfo{volume}{89}},
  \bibinfo{pages}{195407} (\bibinfo{year}{2014}).

\bibitem{Tosi}
\bibinfo{author}{Tosi, L.} \emph{et~al.}
\newblock \bibinfo{title}{Spin-orbit splitting of {Andreev} states revealed by
  microwave spectroscopy}.
\newblock \emph{\bibinfo{journal}{Phys. Rev. X}} \textbf{\bibinfo{volume}{9}},
  \bibinfo{pages}{011010} (\bibinfo{year}{2019}).

\bibitem{Akhmerov}
\bibinfo{author}{Rosdahl, T.~O.}, \bibinfo{author}{Vuik, A.},
  \bibinfo{author}{Kjaergaard, M.} \& \bibinfo{author}{Akhmerov, A.~R.}
\newblock \bibinfo{title}{{Andreev} rectifier: A nonlocal conductance signature
  of topological phase transitions}.
\newblock \emph{\bibinfo{journal}{Phys. Rev. B}} \textbf{\bibinfo{volume}{97}},
  \bibinfo{pages}{045421} (\bibinfo{year}{2018}).

\bibitem{DanonNonlocal}
\bibinfo{author}{Danon, J.} \emph{et~al.}
\newblock \bibinfo{title}{Nonlocal conductance spectroscopy of {Andreev} bound
  states: Symmetry relations and bcs charges}.
\newblock \emph{\bibinfo{journal}{Phys. Rev. Lett.}}
  \textbf{\bibinfo{volume}{124}}, \bibinfo{pages}{036801}
  (\bibinfo{year}{2020}).

\bibitem{MenardNonlocal}
\bibinfo{author}{M\'enard, G.~C.} \emph{et~al.}
\newblock \bibinfo{title}{Conductance-matrix symmetries of a three-terminal
  hybrid device}.
\newblock \emph{\bibinfo{journal}{Phys. Rev. Lett.}}
  \textbf{\bibinfo{volume}{124}}, \bibinfo{pages}{036802}
  (\bibinfo{year}{2020}).

\bibitem{PugliaNonlocal}
\bibinfo{author}{Puglia, D.} \emph{et~al.}
\newblock \bibinfo{title}{Closing of the induced gap in a hybrid
  superconductor-semiconductor nanowire}.
\newblock \emph{\bibinfo{journal}{Phys. Rev. B}}
  \textbf{\bibinfo{volume}{103}}, \bibinfo{pages}{235201}
  (\bibinfo{year}{2021}).

\bibitem{Sarma3T}
\bibinfo{author}{Pan, H.}, \bibinfo{author}{Sau, J.~D.} \&
  \bibinfo{author}{Das~Sarma, S.}
\newblock \bibinfo{title}{Three-terminal nonlocal conductance in {Majorana}
  nanowires: Distinguishing topological and trivial in realistic systems with
  disorder and inhomogeneous potential}.
\newblock \emph{\bibinfo{journal}{Phys. Rev. B}}
  \textbf{\bibinfo{volume}{103}}, \bibinfo{pages}{014513}
  (\bibinfo{year}{2021}).

\bibitem{DrachmannAnodic}
\bibinfo{author}{Drachmann, A. C.~C.} \emph{et~al.}
\newblock \bibinfo{title}{Anodic oxidation of epitaxial
  superconductor-semiconductor hybrids}.
\newblock \emph{\bibinfo{journal}{Phys. Rev. Materials}}
  \textbf{\bibinfo{volume}{5}}, \bibinfo{pages}{013805} (\bibinfo{year}{2021}).

\bibitem{wimmer_algorithm_2012}
\bibinfo{author}{Wimmer, M.}
\newblock \bibinfo{title}{Algorithm 923: {{Efficient Numerical Computation}} of
  the {{Pfaffian}} for {{Dense}} and {{Banded Skew}}-{{Symmetric Matrices}}}.
\newblock \emph{\bibinfo{journal}{ACM Transactions on Mathematical Software}}
  \textbf{\bibinfo{volume}{38}}, \bibinfo{pages}{30:1--30:17}
  (\bibinfo{year}{2012}).

\bibitem{groth_kwant_2014}
\bibinfo{author}{Groth, C.~W.}, \bibinfo{author}{Wimmer, M.},
  \bibinfo{author}{Akhmerov, A.~R.} \& \bibinfo{author}{Waintal, X.}
\newblock \bibinfo{title}{Kwant: A software package for quantum transport}.
\newblock \emph{\bibinfo{journal}{New Journal of Physics}}
  \textbf{\bibinfo{volume}{16}}, \bibinfo{pages}{063065}
  (\bibinfo{year}{2014}).

\bibitem{altland_nonstandard_1997}
\bibinfo{author}{Altland, A.} \& \bibinfo{author}{Zirnbauer, M.~R.}
\newblock \bibinfo{title}{Nonstandard symmetry classes in mesoscopic
  normal-superconducting hybrid structures}.
\newblock \emph{\bibinfo{journal}{Phys. Rev. B}} \textbf{\bibinfo{volume}{55}},
  \bibinfo{pages}{1142--1161} (\bibinfo{year}{1997}).

\bibitem{schnyder_classification_2008}
\bibinfo{author}{Schnyder, A.~P.}, \bibinfo{author}{Ryu, S.},
  \bibinfo{author}{Furusaki, A.} \& \bibinfo{author}{Ludwig, A. W.~W.}
\newblock \bibinfo{title}{Classification of topological insulators and
  superconductors in three spatial dimensions}.
\newblock \emph{\bibinfo{journal}{Phys. Rev. B}} \textbf{\bibinfo{volume}{78}},
  \bibinfo{pages}{195125} (\bibinfo{year}{2008}).

\bibitem{SmokingGun}
\bibinfo{author}{Das~Sarma, S.}, \bibinfo{author}{Sau, J.~D.} \&
  \bibinfo{author}{Stanescu, T.~D.}
\newblock \bibinfo{title}{Splitting of the zero-bias conductance peak as
  smoking gun evidence for the existence of the {Majorana} mode in a
  superconductor-semiconductor nanowire}.
\newblock \emph{\bibinfo{journal}{Phys. Rev. B}} \textbf{\bibinfo{volume}{86}},
  \bibinfo{pages}{220506} (\bibinfo{year}{2012}).

\end{thebibliography}
\vspace{0.5cm}
\begin{center}
{\bf Acknowledgements}     
\end{center}

We thank Geoff Gardener and Sergei Gronin for contributions to materials design and growth. We thank Andrey Antipov, Roman Lutchyn, Chetan Nayak, and Ivan Sadovskyy for useful discussions, in particular for pointing out the importance of orbital contributions in the model. We thank Karsten Flensberg, Antonio Fornieri, Raquel Queiroz, Noam Schiller, and Saulius Vaitiek\.enas for insightful discussions. We acknowledge support from the Danish National Research Foundation, a research grant (Project 43951) from VILLUM FONDEN, the ERC under the Horizon 2020 Research and Innovation programme (LEGOTOP No. 788715 and HQMAT No. 817799), the DFG (CRC/Transregio 183, EI 519/7-1), the BSF and NSF (2018643), the ISF Quantum Science and Technology (2074/19), and a research grant from Irving and Cherna Moskowitz. H.R.W. and M.R.L. thank the Weizmann Institute of Science for hospitality via the Yutchun Program during the initial phase of the work.

\clearpage

\appendix

\onecolumngrid

\begin{center}
{\bf SUPPLEMENTARY MATERIAL}
\end{center}

\twocolumngrid

\section{Methods}
{\bf Wafer structure:} The wafer structure used in this work consists of an InAs two-dimensional quantum well in epitaxial contact with Al. The wafer was grown on an insulating InP substrate by molecular beam epitaxy comprising
a 100-nm-thick In$_{0.52}$Al$_{0.48}$As matched buffer, a 1$~\mu$m thick step-graded buffer realized with alloy steps from In$_{0.52}$Al$_{0.48}$As to In$_{0.89}$Al$_{0.11}$As (20 steps, 50 nm/step),a 58 nm In$_{0.82}$Al$_{0.18}$As layer, a 4 nm In$_{0.75}$Ga$_{0.25}$As bottom barrier, a 7 nm InAs quantum well, a 10 nm In$_{0.75}$Ga$_{0.25}$As top barrier, two monolayers of GaAs and a 7 nm film of epitaxially grown Al. The top Al layer was grown in the same molecular beam epitaxy chamber used for the rest of the growth, without breaking the vacuum. 

Hall effect measurements were performed in Hall bar devices with Al etched away (see Fig.~S\ref{suppFigS12} for Hall effect measurements). The Hall bar was covered with the same dielectric material as used in the Josephson junction experiments, grown under nominally identical conditions and of the same thickness. A peak electron mobility $\mu= 43,000$~ cm$^2$/Vs was observed at a carrier density of $n=8 \times 10^{11}$~cm$^{-2}$, corresponding to a peak mean free path of  $l_e\sim$~600 nm at top gate voltage $V_{\rm TG}=-0.8$~V. In the junction experiments, we typically use $V_{\rm sc}=- 3$~V to control density under the superconducting leads and $V_1=0-0.1$~V to control density in the barrier region. Different geometries and lateral gate coupling makes it difficult to compare these voltages directly. To get a rough idea, however, taking $l_e$ to be around 600 nm in the junction yields quasi-ballistic motion along the junction, $l \sim 3 l_e$, and ballistic motion across the junction, $w_n \sim l_e/6$.

Transport characterization of a large-area Hall bar with Al in place yielded a critical field of 2.5~T~\cite{fornieri,DrachmannAnodic} for the parent Al layer, considerably larger than field where the gap closure occurs,  $\sim$~0.5~T.   

{\bf Device fabrication:} Devices were fabricated using conventional electron beam lithography. Devices on the same chip were electrically isolated from each other using a self-aligned mesa etch process, first by removing Al using Transene D wet etch, followed by a wet etch in H$_2$O:C$_6$H$_8$O$_7$:H$_3$PO$_4$:H$_2$O$_2$ (220:55:3:3) to remove the semiconductor to a depth of $\sim$ 300 nm. Next,  Al  was selectively removed leaving the Josephson junction and flux loop. A 15 nm thick layer of HfO$_2$ grown at 90$^\circ$C using atomic layer deposition (ALD) was used as the gate dielectric. Gates were defined using electron beam lithography followed by e-beam evaporation of Ti/Au layers of thicknesses (5 nm/20 nm) for finer structures and (5 nm/350 nm) for the bonding pads. The bottom Ti/Au ohmic contact was formed by etching away a U-shaped trench in the mesa and then contacting the InAs 2DEG. An additional HfO$_2$ layer deposited by ALD and subsequent lift-off was used to isolate the Ti/Au ohmic from the superconducting loop and mesa.

{\bf Electrical transport measurements:} Electrical transport measurements were performed in an Oxford Triton dilution refrigerator at a base temperature of 20~mK using conventional low-frequency AC lock-in techniques at 31.5~Hz excitation frequency, an AC excitation amplitude of 3~$\mu$V  and a variable DC voltage $V_{\rm SD}$ for bias spectroscopy. The current through the device was recorded using a low-impedance current-to-voltage converter that was attached to the ohmic contact connected to the superconducting loop. For measuring the third harmonic of the current, a higher AC excitation amplitude of 10~$\mu$V was used. Magnetic field to the sample was applied using a three-axis ($B_x$,~$B_y$,~$B_z$)=(1T,~1T,~6T) vector magnet.

We fabricated 32 devices, of which 10 were measured. We summarize the behavior of these devices. We also estimate the probability $P_{\rm Z,T(B)}$, of observing a ZBCP, which is defined as the percentage of operable $V_1$ gate space that shows stable ZBCPs at the top (bottom) end. See Fig.~S\ref{suppFigS11} for example.

Device 1: Gap reopening with ZBCP at top end, bottom QPC did not work. $P_{\rm Z,T} \sim 15\%$. 

Device 2: Gap reopening at both ends, ZBCP only at the top end. $P_{\rm Z,T} \sim 30\%$.

Device 3: Gap reopening and stable ZBCP at both ends. $P_{\rm Z,T} \sim 10\%$, $P_{\rm Z,B} \sim 5\%$. 

Device 4: Gap reopening and stable ZBCP at both ends. $P_{\rm Z,T} \sim 10\%$, $P_{\rm Z,B} \sim 10\%$. 

Device 5: Gap reopening and stable ZBCP at both ends. $P_{\rm Z,T} \sim 5\%$, $P_{\rm Z,B} \sim 5\%$. 

Device 6: Gap reopening at both ends. Stable ZBCP at top end. ZBCP at bottom end oscillated as a function of in-plane magnetic field. $P_{\rm Z,T} \sim 5\%$, $P_{\rm Z,B} \sim 5\%$. 

Device 7: Gap reopening on both ends. Soft gap at low fields at both ends.

Device 8: Spectroscopy possible at both ends, however induced superconducting gap at the bottom end collapsed at $B_\parallel \sim 150$~mT.

Device 9: Spectroscopy not possible at bottom end.

Device 10: No detectable superconducting gap on either end.

Theoretical simulation: $P_{\rm Z,T}=P_{\rm Z,B}\sim$ 9--18\%.

{\bf Magnetic field alignment:} The sample is oriented with respect to the vector magnet such that $B_x$ of the magnet is nominally along $B_\perp$, the field in the direction perpendicular to the plane of the wafer [Fig.~\ref{fig01}(c)] and $B_z$ of the magnet is nominally parallel to $B_\parallel$ the field direction along the SN interfaces [Fig.~\ref{fig01}(c)]. However, sample misalignment causes the magnet $B_z$ to have a small contribution to $B_\perp$, which controls the flux through the superconducting loop. At non-zero $B_z$, it is therefore necessary to identify the proportional amount of $B_x$  that results in constant flux through the loop. At zero $B_z$, the value of $B_x$ at which the superconducting gap is maximised corresponds to zero and multiples of $\Phi_0$, while distinct phase slips appear at odd multiples of $\Phi_0/2$. This allows us to calibrate the flux through the device at zero $B_z$. At finite $B_z$, the superconducting gap acquires a phase-asymmetric dispersion, and the maxima of the gap cannot be used to track lines of constant flux. Instead, we use the phase slips to identify lines of constant flux through the device. This allows us to define magnetic fields $B_\parallel$ and $B_\perp$ that compensate for the finite tilt of the sample.

{\bf Estimation of flux loop inductance:} The inductance of the superconducting loop is a combination of the geometric inductance and the kinetic inductance of the thin Al layer, and is dominated by the latter~\cite{NicheleRelating}. We estimate the geometric inductance of the loop as $L_G\sim 2.5$~pH. The kinetic inductance of a thin superconductor is proportional to its sheet resistivity and is given as
\begin{equation}
L_K=\frac{l_s}{w_s} \frac{h}{2\pi^2 e} \frac{R_{\square}}{\Delta},
\end{equation}
where $l_s$ and $w_s$ are the length and width of the superconducting stripe defining the superconducting loop including the meanders that are part of the superconducting leads, and $R_{\square}$ is the normal-state sheet resistivity of the Al/InAs layer. $l_s/w_s \sim 240$ in our device and has two contributions, $l_1/w_1\sim 40$ for the U-shaped part of the loop and $l_2/w_2 \sim 200$ arising from the meandering part of the superconducting leads. The sheet resistance $R_{\square}$ in our material is measured as $\sim$6$~\Omega$ in the normal state of a large area Al Hall bar~\cite{DrachmannAnodic} and the superconducting gap $\Delta \sim 200~\mu$eV. This leads to a kinetic inductance $L_K \sim$~1.5 nH, and total inductance $L\sim L_K\sim$~1.5 nH. In our numerical simulations of the flux dependence of Andreev bound state spectrum, we find that $L \sim 2$~nH qualitatively reproduces the features observed in the measured subgap spectra (see Fig.~\ref{fig04}). 

{\bf Model:} To model our device, we use an extension of the Hamiltonian proposed in Refs.~\cite{Pientka,HellFlensberg} to account for finite thickness and include orbital effects. The model is based on two layers of a two-dimensional semiconductor with Rashba spin-orbit coupling. We consider a rectangular device, with the rectangle divided into three parts by width: normal region in the middle with width $w_n$, and superconducting regions on two sides, each of width $w$. In the Nambu basis $(\psi_{\uparrow},\psi_{\downarrow},\psi_{\downarrow}^\dagger,-\psi_{\uparrow}^\dagger)$, the Bogoliubov--de Gennes Hamiltonian is given by:

\begin{equation}\label{eq:Hamiltonian}
\begin{aligned}
    H&=\left[-\frac{\partial^2_x+\partial^2_y}{2m^*}-t_{\perp}\nu_x-\mu(y)+i\alpha(z)\left(\partial_x\sigma_y-\partial_y\sigma_x\right)\right]\tau_z \\
    &+\frac{g(y)\mu_{\rm B} B_\parallel}{2}\sigma_x+\Delta(y,z)\tau^++\Delta^*(y,z)\tau^-,
\end{aligned}
\end{equation}
where $\sigma,\tau,\nu$ are Pauli matrices acting in spin, electron-hole, and layer basis, respectively. Here $m^*$ is the effective mass of electrons in the semiconductor, $\alpha(z)$ is the layer-dependent Rashba spin-orbit coupling strength, $t_{\perp}$ is the inter-layer hopping amplitude, $B_\parallel$ is the magnetic field applied along the junction, and $\mu_{\rm B}$ is the Bohr magneton. The g-factor $g(y)$ is different for the normal and superconducting regions, such that
\begin{equation}
    g(y)=\left\{ \begin{array}{ll}
g_{\rm N} & \,|y|<\frac{w_n}{2}\\
g_{\rm S} & \, \frac{w_n}{2}<|y|<w+\frac{w_n}{2}.
\end{array} \right.
\end{equation}
Similarly, the chemical potential $\mu$ takes the values $\mu_{\rm N}$ in the normal region and $\mu_{\rm S}$ in the superconducting region.
In the last two terms, $\Delta(y)$ is the superconducting pairing potential which is non-zero only in the superconducting region:
\begin{equation}
    \Delta(y)=\left\{ \begin{array}{ll}
0 & \, |y|<\frac{w_n}{2}\\
\Delta e^{i\phi/2} & \, \frac{w_n}{2}<y<w+\frac{w_n}{2}\\
\Delta e^{-i\phi/2} & \, -w-\frac{w_n}{2} < y < -\frac{w_n}{2}.
\end{array} \right.
\end{equation}

To model the finite thickness of the system, thereby accounting for the orbital effects of the in-plane magnetic field, we utilize the two-layer structure. Hopping between the two layers is described by the amplitude $t_{\perp}$. The orbital effect enters as a vector potential $\vec{A}=B_{\parallel}y\hat{z}$, where $z$ is the out-of-plane direction; the vector potential is incorporated into the tight-binding Hamiltonian as a complex amplitude with the Peierls substitution~\cite{peierls_zur_1933}. Furthermore, the parallel magnetic field induces linear phase growth along the junction's cross section~\cite{tinkham_introduction_2004}, which is modeled as an additional modulation $\propto B_{\parallel}yz$ to the superconducting phase. We note that another possibility of modeling the phase evolution is assuming the proximity effect is only present at the top layer and calibrating $\Delta$ accordingly. The basis of this approach is integrating out the proximitizing superconductor's degrees of freedom, and it yields very similar results to the ones we report here.
In reality a detailed simulation of the system is more involved. It should include the effect of disorder in the Al and InAs layers, and consider a well with finite thickness in the $z$ direction. We should therefore treat the Hamiltonian introduced in Eq.~\eqref{eq:Hamiltonian} as a phenomenological model that,  with a proper choice of effective parameters, reproduces qualitatively the experimental observations. 

For our numerical calculations, we discretize the Hamiltonian to a tight-binding model on a square lattice of spacing $a=10$~nm. Simulations are performed with the following parameters: $m^* = 0.026\,m_e$, $\Delta=140~\mu$eV, $t_{\perp}=10$~meV, $l=4~\mu$m, $w_n=100$~nm, $w = 200$~nm, $w_z=10$~nm, $\mu_{\rm SC}=3.6$~meV, $\mu_{\rm N}=3.3$~meV. The g-factors are taken to be $g_{\rm N}=8$ and $g_{\rm S}=4$. We use the leads spectroscopy measurements [Fig.~S\ref{suppFigS1}] to match $A_{\rm eff}$ (the effective cross section for the field-induced superconducting phase gradient), $\mu_{\rm SC}$, and the difference in spin-orbit coupling between the two layers. We obtained $A_{\rm eff}=0.4(2w+w_n)w_z$, $\mu_{\rm SC}=3.6$~meV, and the spin-orbit coupling constants $\alpha(0) = 15$~meV~nm, $\alpha(1)=-\alpha(0)/4$. Pfaffians were computed using the pfpack software package~\cite{wimmer_algorithm_2012}. Some of the preliminary simulations were performed using the Kwant software package~\cite{groth_kwant_2014}.

We further introduce the effect of finite loop inductance to simulate the flux jumps observed in the experimental phase spectra by establishing the relation between the external flux ($\Phi$) penetrating the device and the phase difference ($\phi$) dropped across the Josephson junction. The spectra of the system obtained as a function of $\phi$ can then be mapped to spectra as a function of the applied flux $\Phi$. Given a phase difference $\phi$, we calculate the ground-state energy $E_{\rm GS}(\phi)$ by summing up the energies of all occupied levels $E<0$. We then calculate the supercurrent at zero temperature, $I(\phi) = -\frac{dE_{\rm GS}}{d\phi}$. In the presence of a finite loop inductance $L$, the external flux $\Phi$ and the phase difference across the Josephson junction $\phi$ are related as $\Phi = (\Phi_0/2\pi) \phi - L I(\phi)$,  where the second term accounts for the magnetic flux dropped across the flux loop when a supercurrent $I(\phi)$ flows through it. For each $\Phi$, several values of $\phi$ may be possible. We use a quasi-static approximation and choose the value of $\phi$ that minimises the total energy $E_{\rm tot}(\phi)=E_{\rm GS}(\phi) + \frac{1}{2}L I^2(\phi)$, where the second term is the magnetic energy stored in the loop. Once the mapping $\Phi \to \phi$ is established, we obtain the energy spectrum as a function of $\Phi$. Here we provide simulations with $L=2$~nH. 

In addition, we examined the effect of disorder by introducing a random potential term $V(x,y)$ into the Hamiltonian. We took $V(x,y)$ to be a random uncorrelated Gaussian variable, $\left\langle V(x,y) V(x',y') \right\rangle = V_0^2 \delta(x-x')\delta(y-y')$. The corresponding tight-binding version of this random potential is a site-dependent random addition to the chemical potential, whose variance $V_{\rm TB}$ is related to the continuum value $V_0$ by $V_{\rm TB}^2 a^2 = V_0^2 = \frac{1}{m^{*}\tau} = \frac{2a^2 t_{\parallel}}{\tau}$, where $\tau$ is the transport lifetime. Therefore, $V_{\rm TB}^2 = 2t_{\parallel}/\tau$. We took $\hbar/\tau=1\,{\rm meV}$ in the region not covered by the superconductor and $\hbar/\tau=0.5\,{\rm meV}$ for the covered region (due to the lower Fermi velocity there). For this intermediate range of $\tau$, roughly consistent with the mean free path from Fig.~\ref{suppFigS12}, disorder may or may not destroy the zero-energy state, depending on the particular disorder realization. When $\tau$ is increased by a factor of 10 (weak disorder) the zero-energy state is almost always observed, while decreasing $\tau$ by a factor of 10 (strong disorder) essentially eliminates the zero-energy state.


\onecolumngrid
\appendix

\makeatletter
\renewcommand{\fnum@figure}{\figurename~S\thefigure}
\makeatother
\setcounter{figure}{0}

\begin{figure*}[htb]
\includegraphics[width=0.9\textwidth]{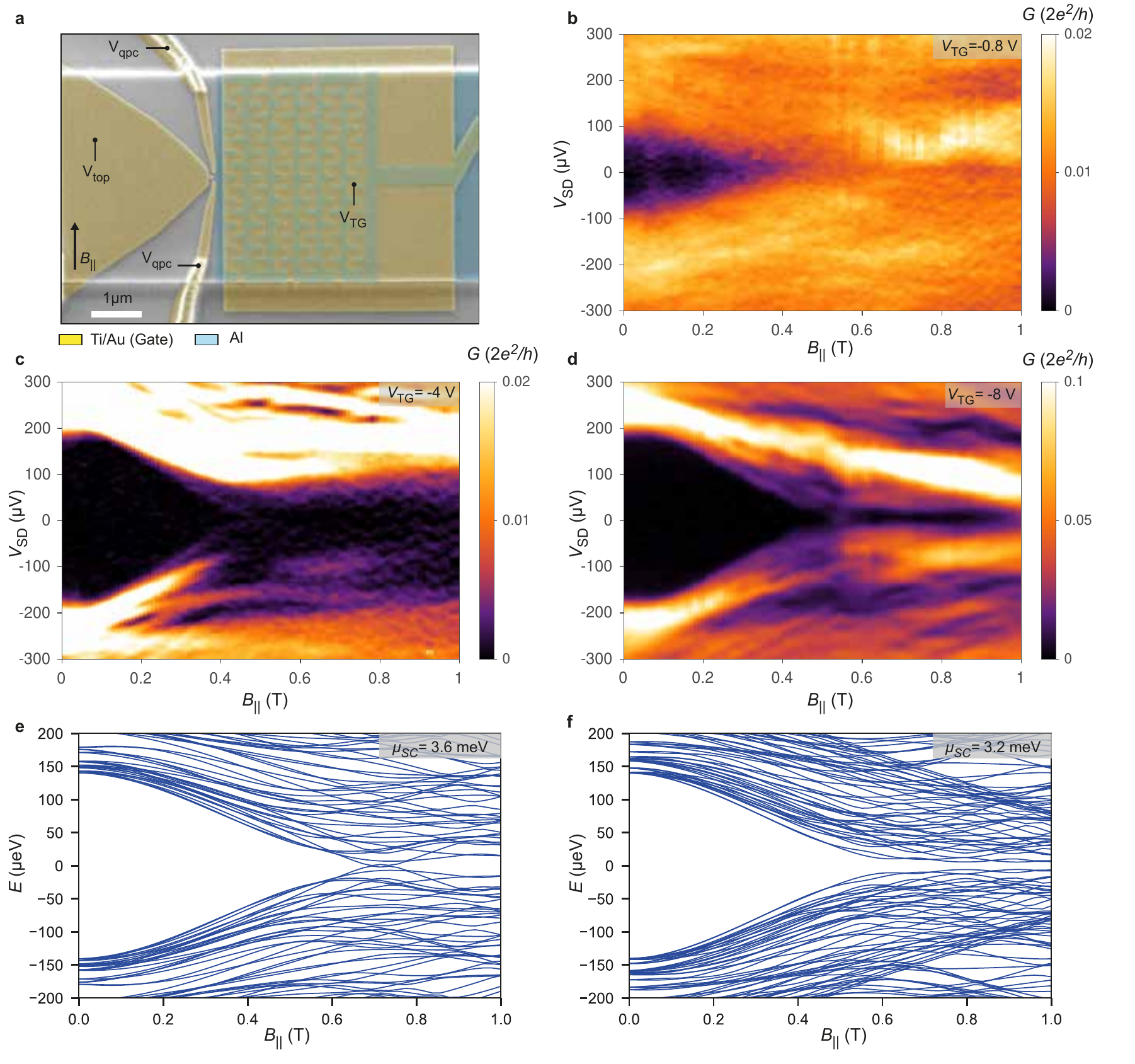}
\caption{\label{suppFigS1} {\bf  Spectroscopy of perforated superconducting leads.} (a)~False-color scanning electron micrograph of a representative device used to study the gap under the superconducting leads. The superconducting lead has meandering perforations of dimensions equivalent to those used in the planar Josephson junction devices. $V_{\rm TG}$ allows gate control of the electron density in the regions where Al has been etched away. Tunnelling spectroscopy is performed using quantum point contacts that are electrostatically defined using a combination of $V_{\rm qpc}$ and $V_{\rm top}$ gate voltages. (b)~Differential conductance $G$ as a function of in-plane magnetic field $B_{\parallel}$ at  $V_{\rm TG}=-0.8$~V displays a soft superconducting gap that collapses at $B_{\parallel} \sim$ 0.3~T. The hardness of the gap is significantly improved by depleting carriers using $V_{\rm TG}$. (c)~At $V_{\rm TG}=-4.0$~V and (d)~$-8.0$~V a hard superconducting gap is obtained that persists until $B_{\parallel} \sim 0.6$~T. Numerical simulations of the spectrum underneath the superconducting leads are shown for (e)~$\mu_{\rm SC}=3.6$~meV and (f)~$\mu_{\rm SC}=3.2$ meV. The parameters of the model are adjusted to match the spectrum of the leads: $A_{\rm eff}$, $\mu_{\rm SC}$, and the difference in spin-orbit coupling between the layers. Simulations for the planar Josephson junction device are performed using the same parameters. }
\end{figure*}

\begin{figure*}[htb]
\includegraphics[width=1\textwidth]{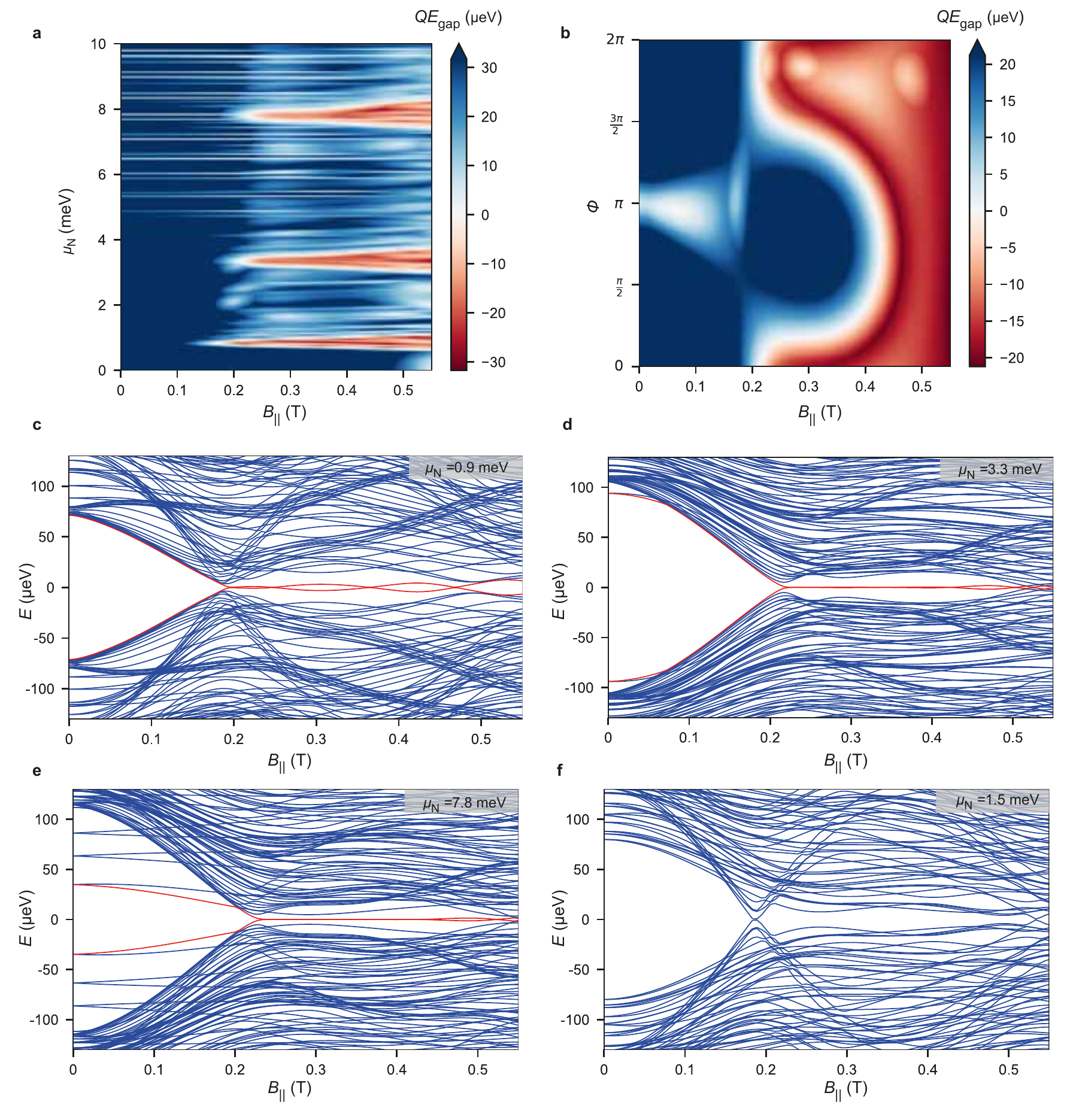} 
\caption{
{\bf Theoretical phase diagrams and spectra.}  (a)~Topological phase diagram in the plane of the parallel magnetic field $B_{\parallel}$ and the chemical potential $\mu_{\rm N}$ in the normal region (at $\phi=0$). The colors indicate the topological invariant ${\cal Q}$, which is $+1$ ($-1$) for the trivial (topological) phase, multiplied by the energy gap. The diagram exhibits appreciable topological regions starting near $B_{\parallel}\sim 0.2$~T. Near closings of the gap at $B_{\parallel}\sim 0.2$~T are almost independent of $\mu_{\rm N}$.  These are not $k=0$ gap closings and are thus not related to a class-D topological phase transition~\cite{altland_nonstandard_1997,schnyder_classification_2008}.
(b)~Topological phase diagram in the $B_{\parallel}$--$\phi$ plane (at $\mu_{\rm N}=3.3$ meV), showing that as a function of $B_{\parallel}$, the system can support a topological phase for all, none, or some values of $\phi$.
The spectra of finite junctions are shown for (c)~$\mu_{\rm N}=0.9$ meV, (d)~$\mu_{\rm N}=3.3$ meV, (e)~$\mu_{\rm N}=7.8$ meV, (f)~$\mu_{\rm N}=1.5$ meV. While (c)--(e) correspond to the three topological regions shown in (a), and therefore support Majorana zero modes (red), the spectrum in (f) does not undergo a topological transition; instead, the gap nearly closes around $B_{\parallel}\sim0.2$~T and then reopens without a zero-energy state.
\label{suppFigS2}}
\end{figure*}

\begin{figure*}[htb]
\includegraphics[width=1\textwidth]{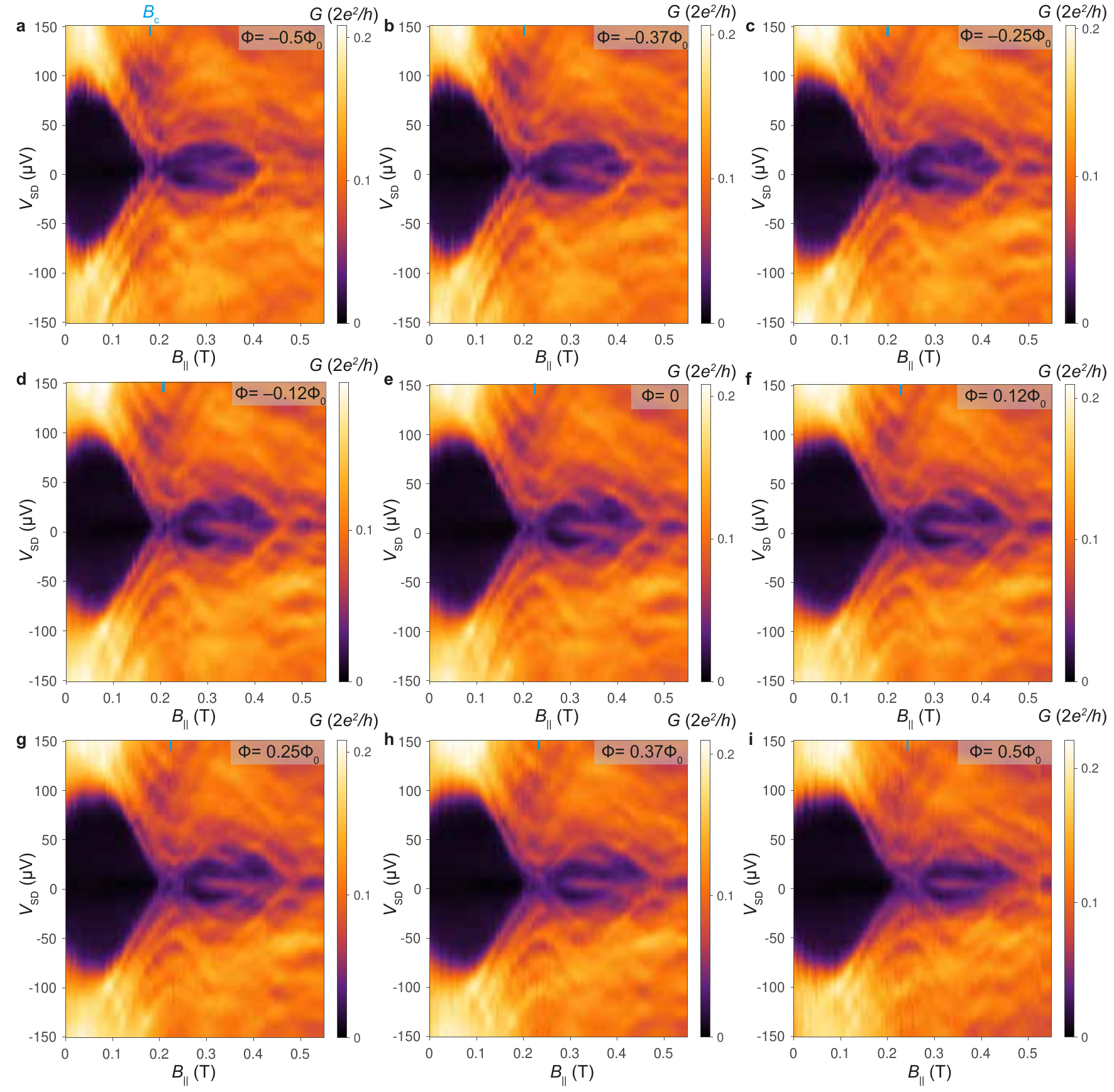} 
\caption{\label{suppFigS3}  {\bf Device 1: Reopening of the gap at different phase biases.} Differential conductance measured as a function of in-plane magnetic field at different values of the flux threading the superconducting loop varying from (a)~$\Phi=-0.5\Phi_0$ to (i)~$\Phi=0.5 \Phi_0$. The value of gap-reopening field $B_c$ shows a variation ($\sim$ 50mT) with the externally imposed magnetic flux $\Phi$. The closing of the gap and the appearance of the zero-bias state remain correlated with the variation of flux. }
\end{figure*}

\begin{figure*}[htb]
\includegraphics[width=1\textwidth]{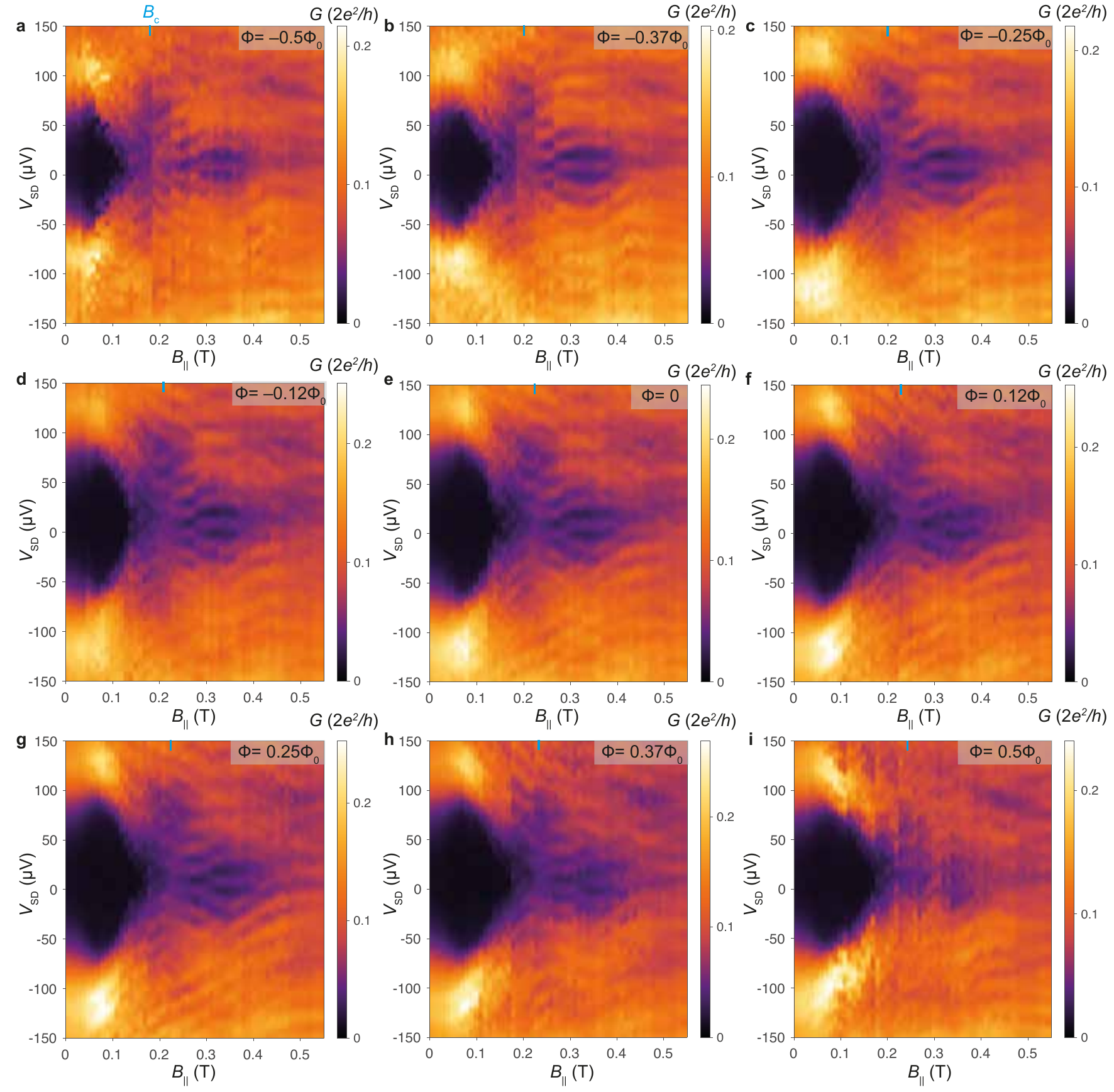} 
\caption{\label{suppFigS4} {\bf Device 2: Reopening of the gap at different phase biases.} Differential conductance measured as a function of in-plane magnetic field at different values of the flux threading the superconducting loop varying from (a)~$\Phi=-0.5\Phi_0$ to (i)~$\Phi=0.5 \Phi_0$. The value of gap-reopening field $B_c$ shows a variation ($\sim$ 40~mT) with the externally imposed magnetic flux $\Phi$. The closing of the gap and the appearance of the zero-bias state remain correlated with the variation of flux.}
\end{figure*}

\begin{figure*}[htb]
\includegraphics[width=1\textwidth]{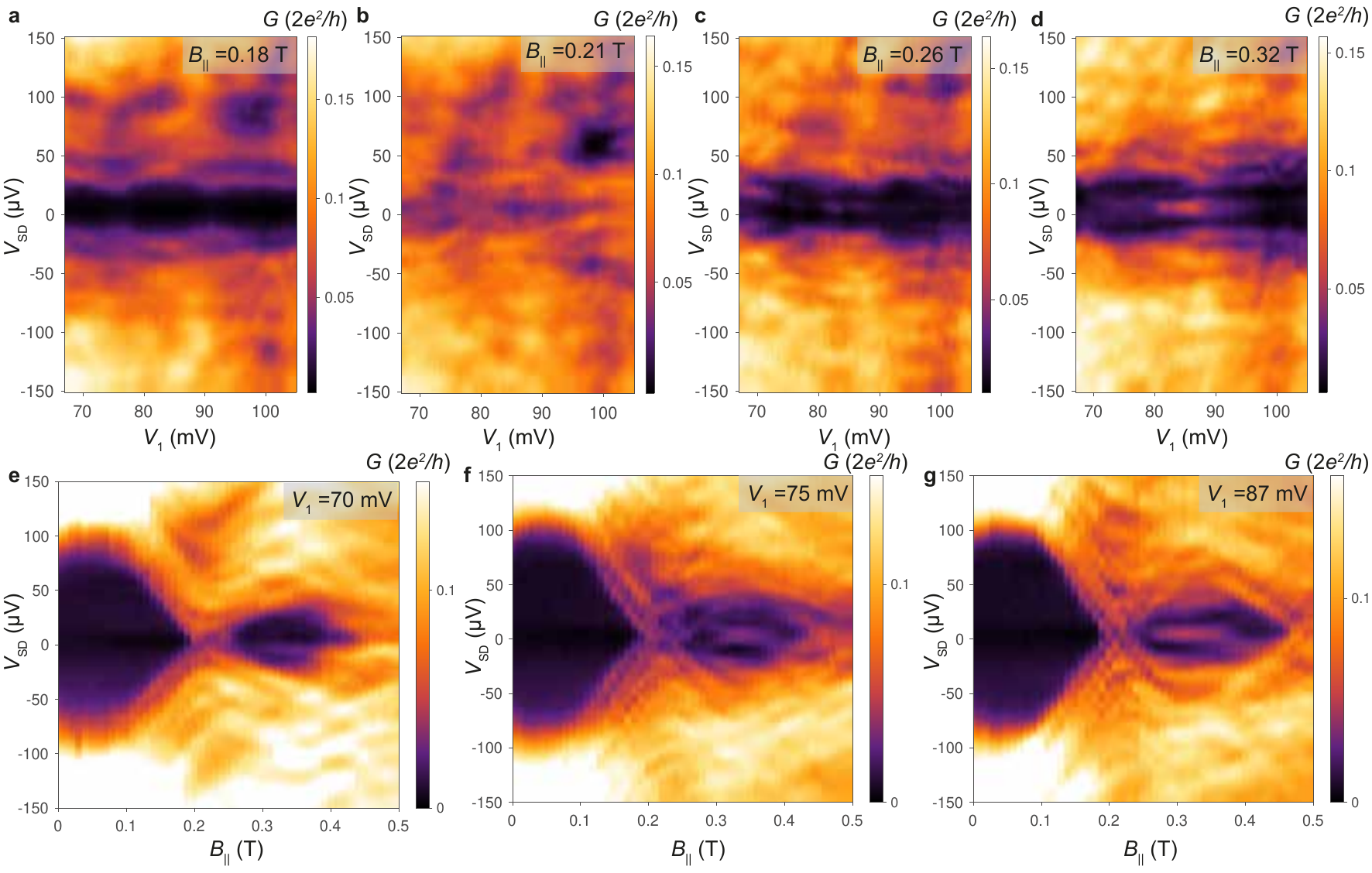} 
\caption{\label{suppFigS5}  {\bf Device 1: Reopening of the gap as a function of chemical potential.} Differential conductance measured as a function of $V_1$ at different values of in-plane magnetic field. (a)~$B_\parallel=0.18$~T,  the spectrum is fully gapped with no low-energy states. (b)~$B_\parallel=0.21$~T, the spectrum becomes gapless. (c)~$B_\parallel=0.26$~T, the gap reopens with sub-gap state formation near zero energy. (d)~$B_\parallel=0.32$~T, the spectrum is fully gapped with a stable zero-bias conductance peak. We also evaluate the field dependence of the spectrum at different values of $V_1$. Differential conductance measured as a function of in-plane magnetic field and $\Phi=0$ at (e)~$V_1=70$~mV, the gap reopens without the formation of a zero-energy state (f)~75~mV, and (g)~87~mV.  The zero-energy state exhibits oscillatory splitting behavior that is reminiscent of field-dependent hybridization of Majorana zero modes in a finite-length 1D topological superconductor~\cite{SmokingGun}.
 }
\end{figure*}

\begin{figure*}[htb]
\includegraphics[width=1\textwidth]{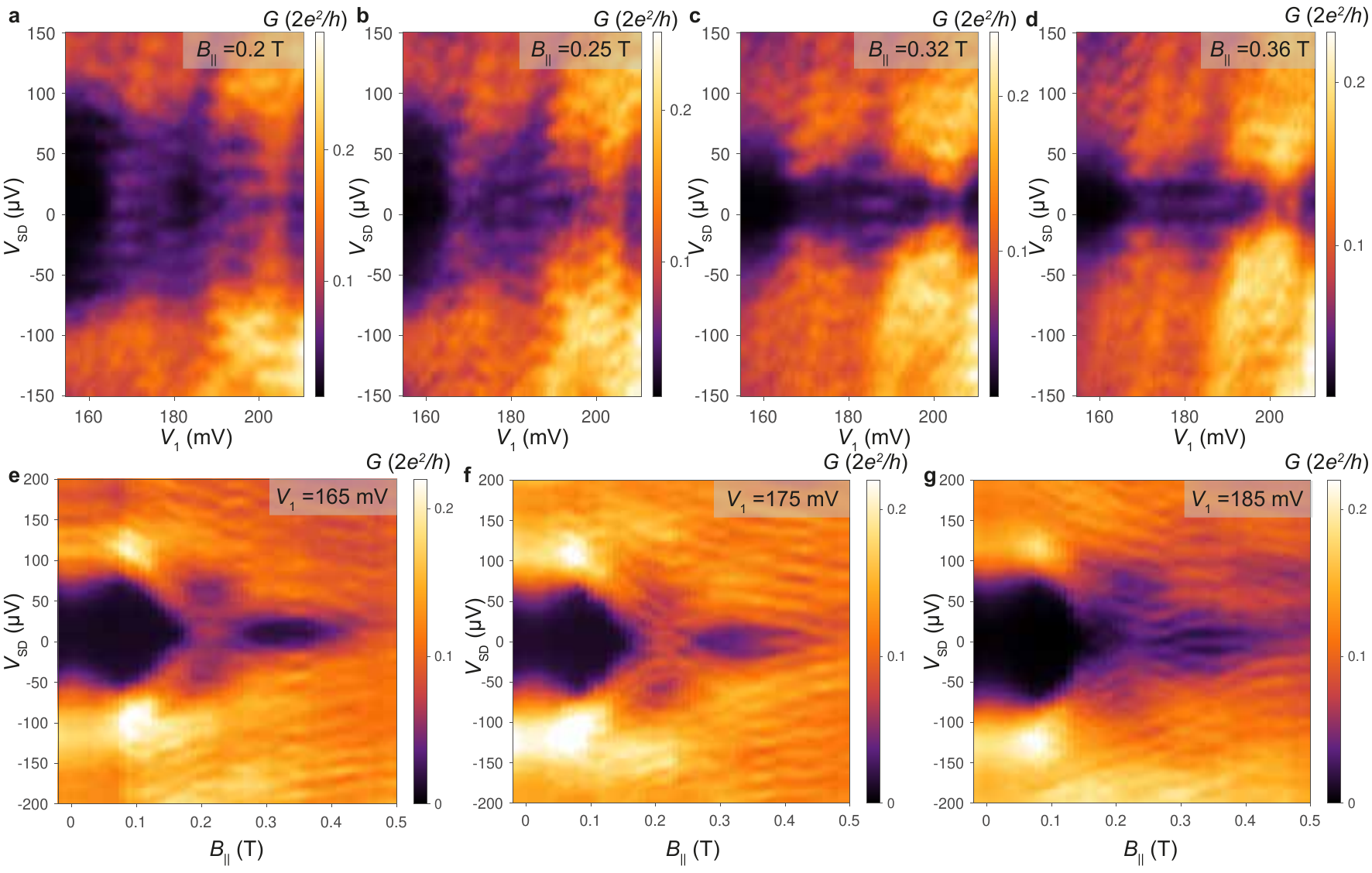} 
\caption{\label{suppFigS6}  {\bf Device 2: Reopening of the gap as a function of chemical potential.} Differential conductance measured as a function of $V_1$ at different values of in-plane magnetic field. (a)~$B_\parallel=0.2$~T,  the spectrum is gapless. (b)~$B_\parallel=0.25$~T, the superconducting gap begins to reopen with zero-bias peak formation. (c)~$B_\parallel=0.32$~T, the gap reopens maximally with a stable zero-bias conductance peak. (d)~$B_\parallel=0.36$~T, the zero-bias conductance peak has a larger span in $V_1$. We inspect the in-plane field dependence of the spectrum at different values of $V_1$. Differential conductance measured as a function of in-plane magnetic field and $\Phi=0$ at (e)~$V_1=165$~mV, we observe the closing and reopening of the superconducting gap without the formation of a zero-bias conductance peak. (f)~$V_1=$175~mV, we observe a zero-bias conductance peak that stabilizes around $B_\parallel=0.35$~T.  (g)~$V_1=$185~mV, the zero-bias conductance peak is stable from $B_\parallel=0.3$~T to $B_\parallel=0.45$~T }
\end{figure*}

\begin{figure*}[htb]
\includegraphics[width=1\textwidth]{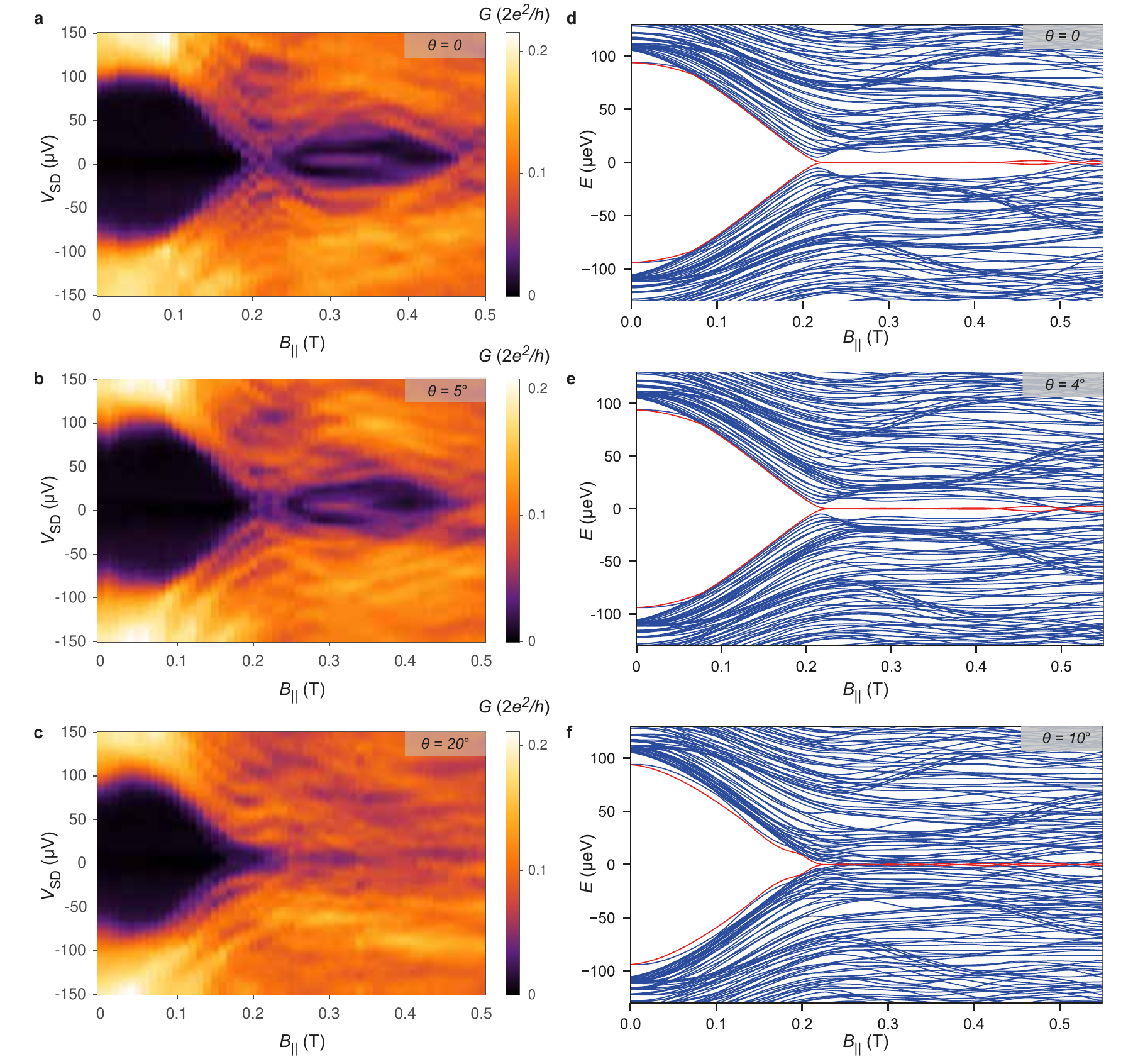} 
\caption{\label{suppFigS7} {\bf Device 1: Tilting of the in-plane magnetic field.}  Differential conductance measured as a function of in-plane magnetic field where a magnetic field of magnitude $B_\parallel$ is applied in the plane of the sample with angle $\theta$ with respect to the axis of the Josephson junction. (a)~At $\theta=0$, we observe reopening of the gap with formation of a zero-bias conductance peak that splits at $B_\parallel=0.38$~T. (b)~At $\theta=5^{\circ}$, the reopening of the gap is suppressed and the zero-bias conductance peak begins to split at a smaller magnetic field ($B_\parallel=0.33$~T) compared to $\theta=0$. (c)~At $\theta=20^{\circ}$, the reopening of the gap is completely suppressed. The critical magnitude of the field for the reopening remains fixed at $B_\parallel$=0.22~T for different values of $\theta$. (d)--(f)~Model spectra in the presence of a tilted in-plane magnetic field for $\theta=0$, $\theta=4^{\circ}$ and $\theta=10^{\circ}$. The tilted field is modeled as a modified Zeeman term $\frac{g\mu_{\rm B}B_{\parallel}}{2}\left(\sigma_x \cos\theta + \sigma_y\sin\theta \right)$, and vector potential, $\vec{A}=B_{\parallel}\left(\hat{z}y\cos\theta+\hat{x}z\sin\theta \right)$, which maintains translational invariance along $x$. }
\end{figure*}

\begin{figure*}[htb]
\includegraphics[width=1\textwidth]{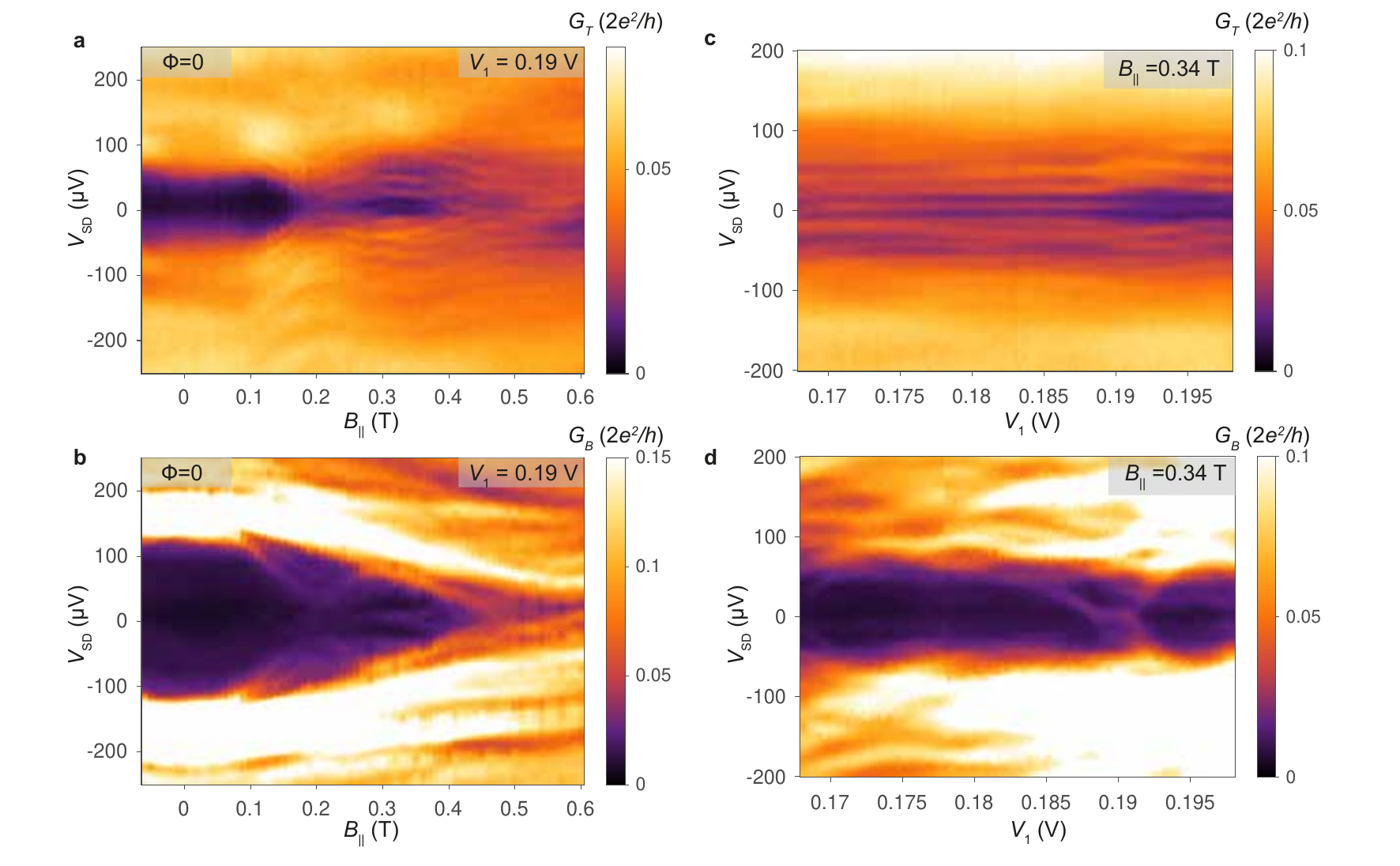}
\caption{\label{suppFigS8}  {\bf Device 3: Concurrent tunnelling spectroscopy at the two ends of the junction.} Differential conductance measured as a function of source-drain bias $V_{\rm SD}$ and in-plane magnetic field $B_{\parallel}$, (a)~$G_{\rm T}$~the top end and (b)~$G_{\rm B}$~at the bottom end. The phase bias is set to $\Phi=0$. Both ends display a closing and reopening of the gap at $B_{\parallel} \sim $~0.2~T. Both ends display formation of subgap states after the reopening of the gap.  (c) and (d) Differential conductance measured as a function of $V_1$ simultaneously at the top and bottom ends of the junction at $B_{\parallel} \sim $~0.34~T. The ZBCP is stable for a larger range of $V_1$ at the top end compared to the bottom end. }
\end{figure*}

\begin{figure*}[htb]
\includegraphics[width=1\textwidth]{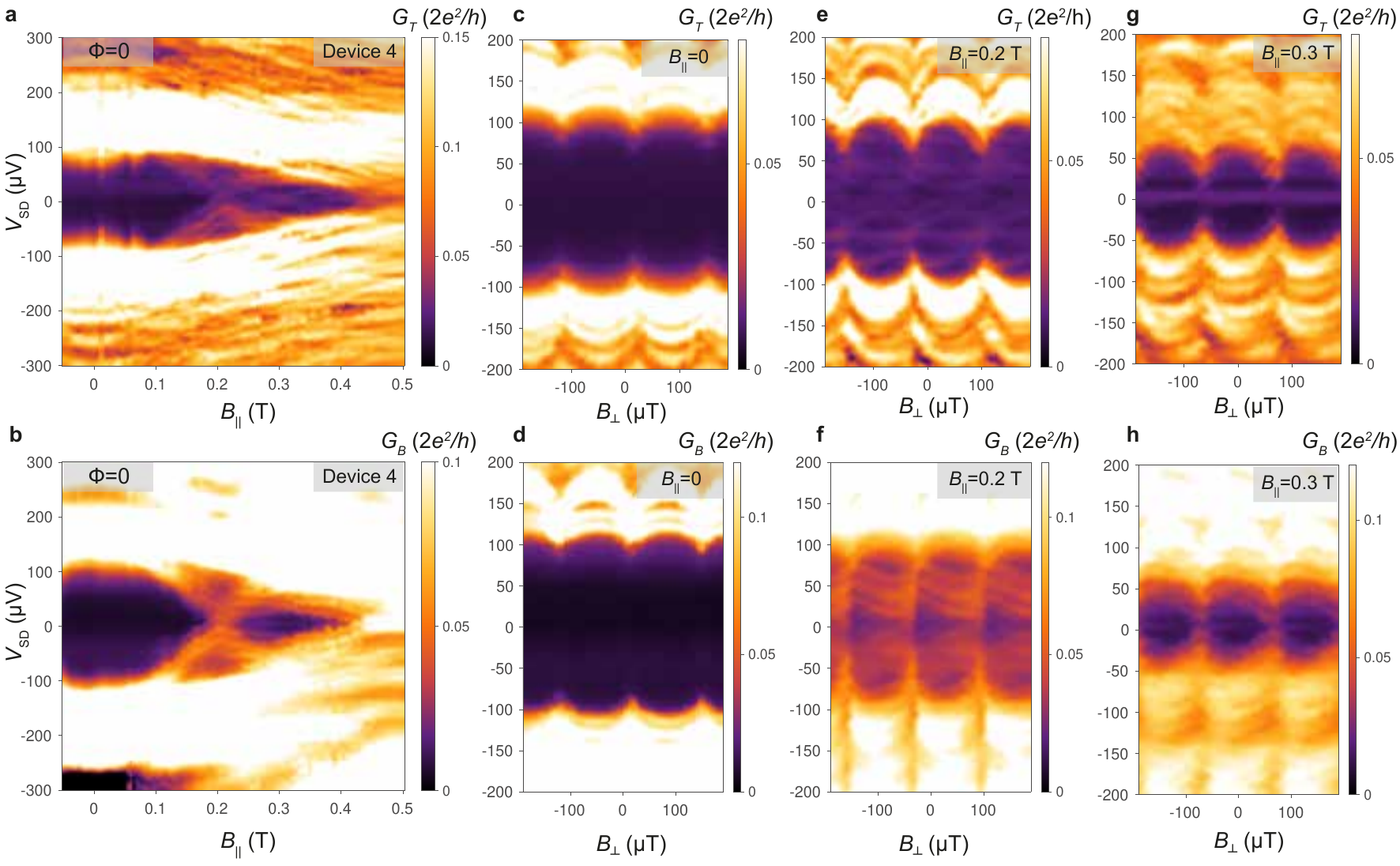} 
\caption{\label{suppFigS9}  {\bf Device 4: Simultaneous tunnelling spectroscopy at the two ends of the junction.} Differential conductance measured as a function of source-drain bias $V_{\rm SD}$ and in-plane magnetic field $B_{\parallel}$. (a)~$G_{\rm T}$~the top end and (b)~$G_{\rm B}$~at the bottom end. The phase bias is set to $\Phi=0$. Both ends display a closing and reopening of the gap at $B_{\parallel} \sim $0.2~T. Both ends display formation of subgap states after the reopening of the gap. At the top end, the subgap state oscillates around zero bias, and forms a zero-bias conductance peak at $B_\parallel=0.3$~T. At the bottom end, a stable zero-bias conductance peak appears after the reopening of the gap. The induced gap collapses simultaneously at $B_\parallel=0.45$~T at both the ends.
Simultaneous differential conductance measured at the top and bottom ends as a function of source-drain bias $V_{\rm SD}$ and out-of-plane magnetic field $B_{\perp}$ for different values of in-plane magnetic field $B_{\parallel}$. (c) and (d) At $B_{\parallel}=$~0, the superconducting gap is modulated periodically at both ends as a function of $B_{\perp}$  with the same periodicity and zero relative phase difference. (e) and (f) At $B_{\parallel}=$~0.2~T,  the spectrum at both ends becomes gapless for all values of $B_{\perp}$  (g) and (h) At $B_{\parallel}=$~0.3~T the superconducting gap reopens. Both the top and bottom ends display a zero-bias conductance peak which is stable with respect to variation of phase. Gate voltages were $V_1=+93$~mV, $V_{\rm SC}=-3.5$~V, $V_{\rm qpc,top}=-0.375$~V, $V_{\rm top}=+0.1$~V, $V_{\rm qpc,bot}=-0.35$~V, $V_{\rm bot}=+0.09$~V, and $V_{\rm loop}=-3.0$~V. }
\end{figure*}

\begin{figure*}[htb]
\includegraphics[width=1\textwidth]{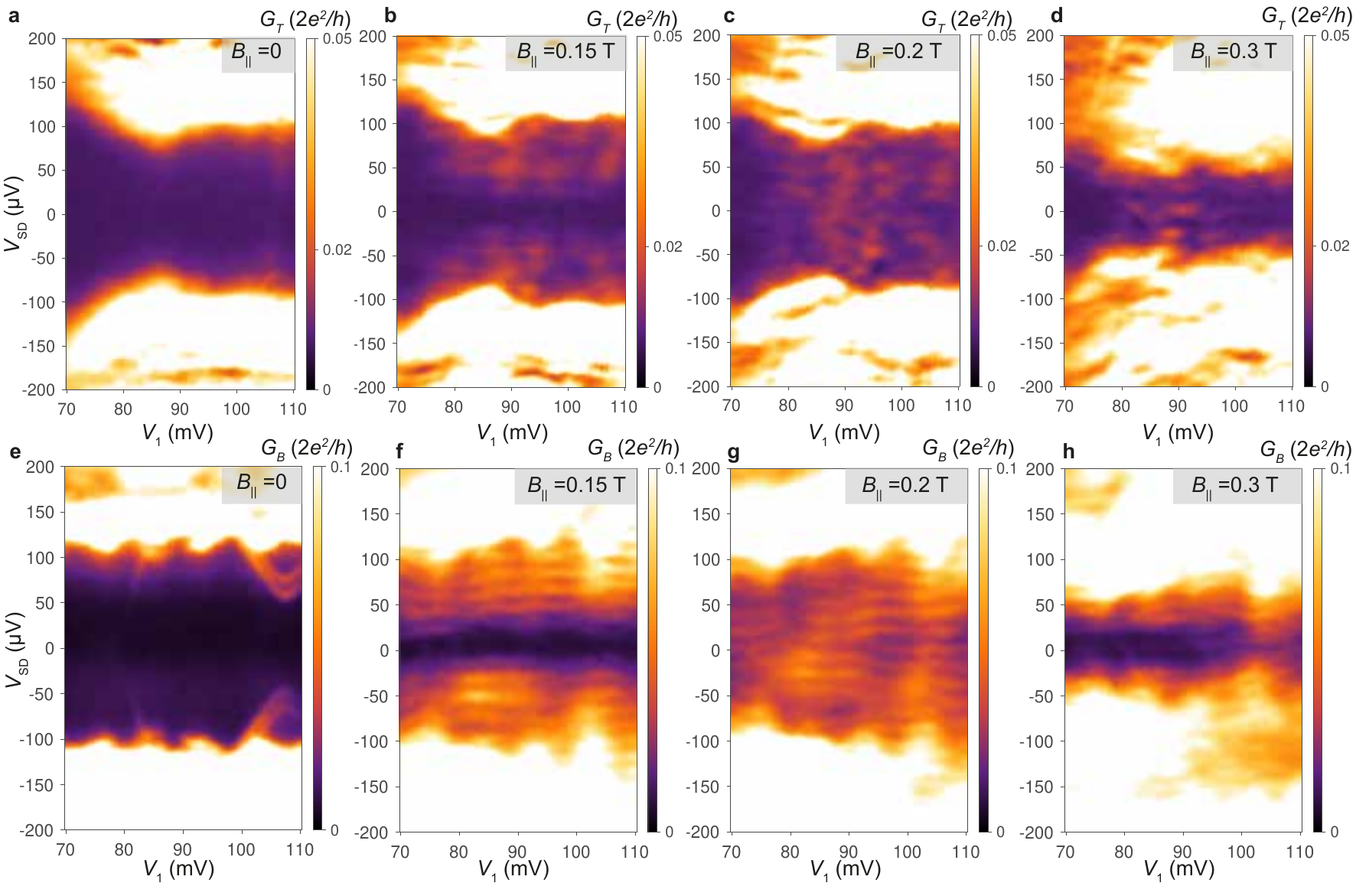} 
\caption{\label{suppFigS10} {\bf Device 4: Concurrent reopening of the gap as a function junction gate at the two ends of the junction.} Differential conductance measured as a function of $V_1$ simultaneously at the top and bottom ends of the junction at different values of in-plane magnetic field. (a, e)~$B_\parallel=0$,  (b, f)~$B_\parallel=0.15$~T, subgap states are lowered in energy but the gap is still finite. (c, g)~$B_\parallel=0.2$~T, the superconducting gap closes at both the top and bottom ends. (d, h)~$B_\parallel=0.3$~T, the gap reopens with the formation of subgap states with zero-bias conductance peaks for certain ranges of $V_1$. The subgap states do not exhibit regions of sustained correlation as a function of $V_1$. }
\end{figure*}

\begin{figure*}[htb]
\includegraphics[width=1\textwidth]{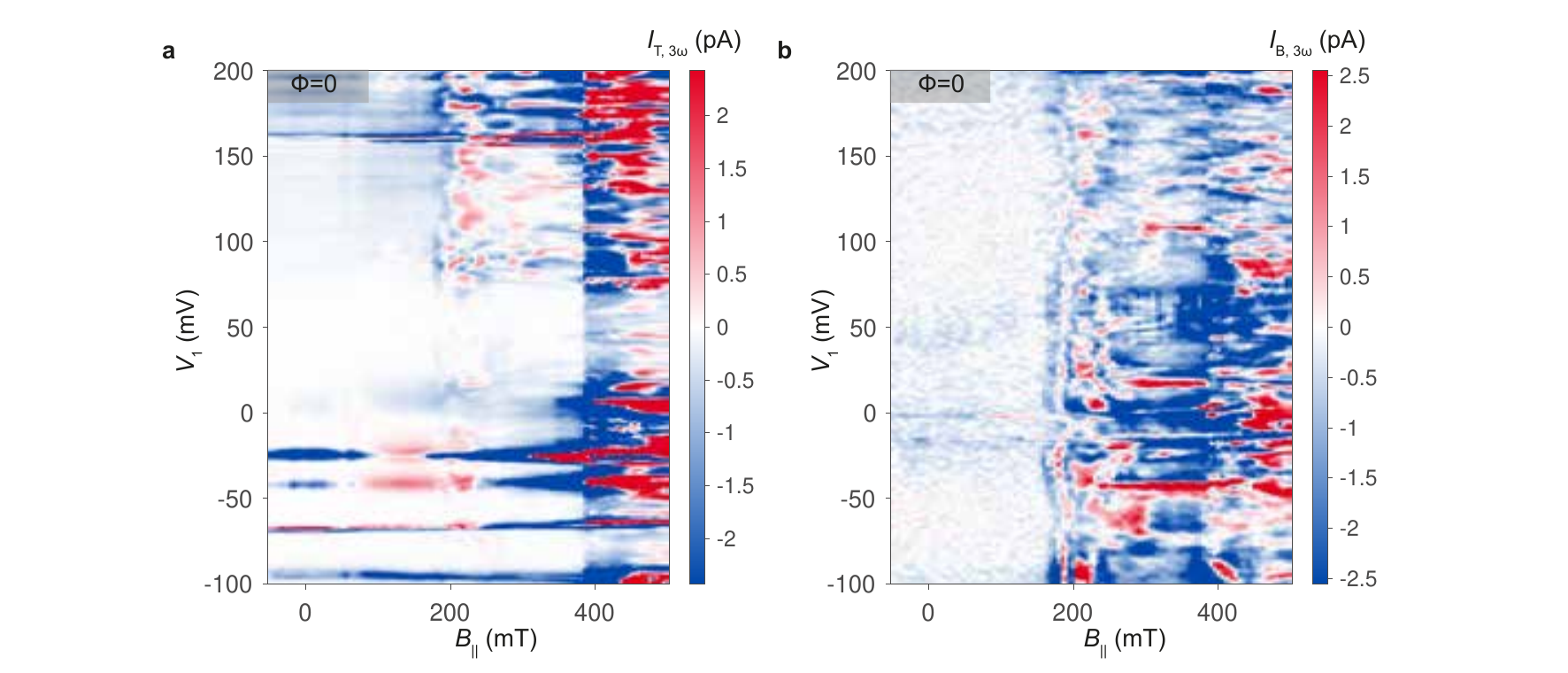}
\caption{\label{suppFigS11} {\bf Device 4: Third harmonic of the current at the two ends.}  Third-harmonic current measured at $V_{\rm SD}=0$ as a function of $V_1$ and in-plane magnetic field $B_{\parallel}$ (a) $I_{\rm T,3\rm \omega}$ at the top and (b) $I_{\rm B,3\rm \omega}$ at bottom of the junction, at phase bias $\Phi=0$. A positive signal indicates a zero-bias conductance peak~\cite{fornieri}. The first closing of the gap produces a region with intermittent positive third-harmonic current at $B_{\parallel} \sim 200$~mT at both ends. At $B_{\parallel}>~$400~mT, the third-harmonic current is positive for a sizable fraction of $V_1$, associated with a reclosing of the gap. For 200~mT~$\leq B_{\parallel} \leq $400~mT, regions of $V_1$ with positive third harmonic correspond to ZBCPs after the gap-reopening. At $B_\parallel=300$~mT, we estimate the percentage of $V_1$ space that produces a positive third harmonic signal as $P_{\rm Z,T} \sim 11\%$ at the top tunnel probe and $P_{\rm Z,B} \sim 13\%$ at the bottom tunnel probe.}
\end{figure*}

\begin{figure*}[htb]
\includegraphics[width=1\textwidth]{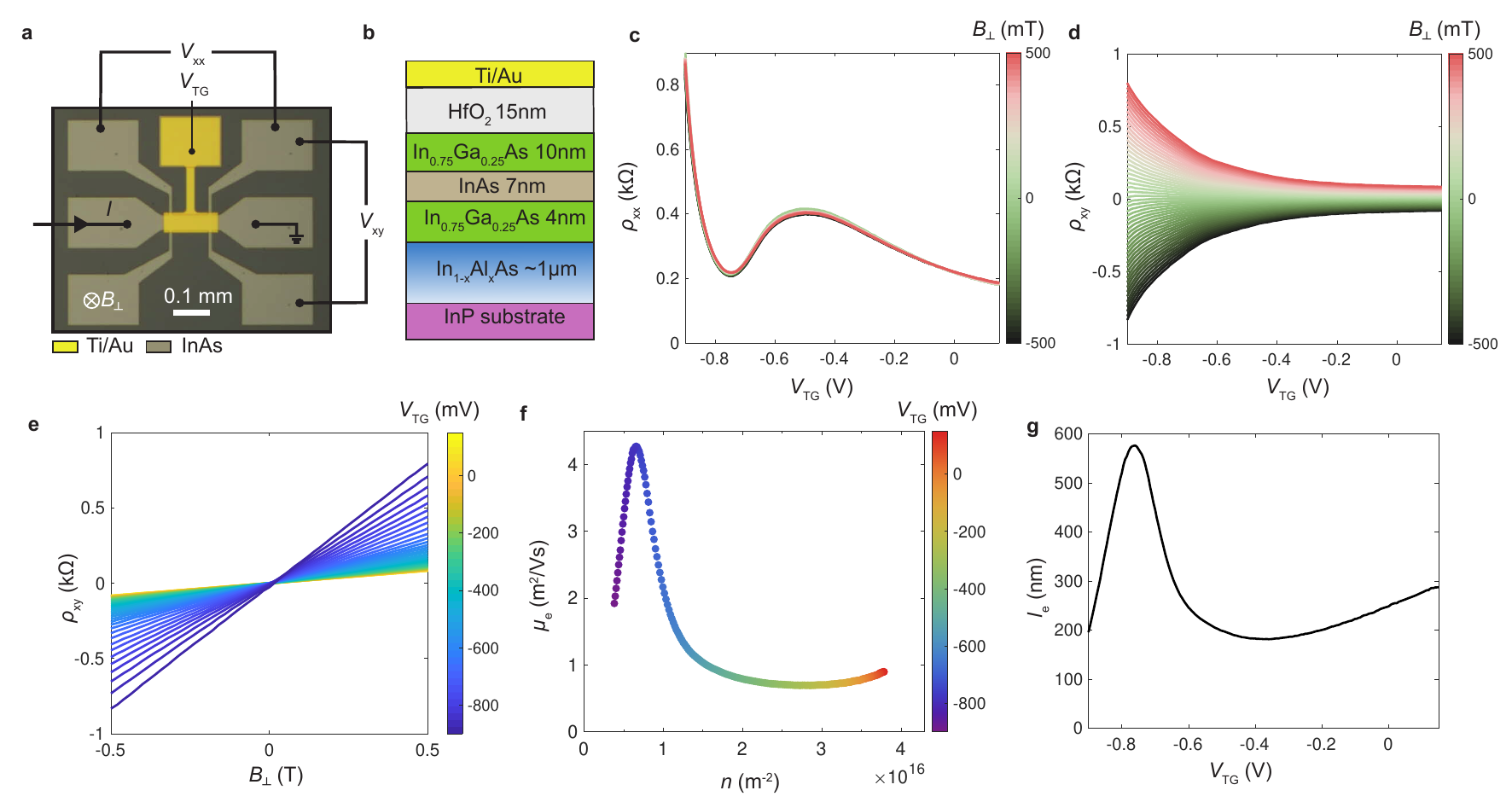}
\caption{\label{suppFigS12} {\bf Hall effect measurement.} (a) Optical micrograph of a Hall-bar device. The Al layer was etched away using Transene D etchant, followed by wet-etching of the heterostructure to define a mesa in the shape of a 6-probe Hall bar. Next, HfO$_2$ dielectric of thickness 15~nm (same as that used in the JJ experiments) was deposited globally on the entire sample. A Ti/Au top gate was then deposited using electron-beam evaporation. For Hall effect measurement, an AC current bias of amplitude $I = 10$~nA, and excitation frequency 166 Hz, was applied to the source terminal of the device with the drain grounded. Longitudinal voltage, $V_{\rm xx}$ and transverse voltage, $V_{\rm xy}$ were measured using two separate separate lock-in amplifiers, as a function of the top-gate voltage, $V_{\rm TG}$. $V_{\rm TG}$ controls the carrier density in the active region. (b) Schematic cross-section of the active region showing the layers of the heterostructure, the dielectric and the Ti/Au gate. (c) Longitudinal sheet resistance $\rho_{\rm xx}=(V_{\rm xx}/I)(W/L)$, where $L/W=2.5$ is the aspect ratio of the active region. (d) Transverse resistance $\rho_{\rm xy}=V_{\rm xy}/I$, measured as a function of $V_{\rm TG}$ at different values of out-of-plane magnetic field $B_{\perp}$. (e) Transverse resistance $\rho_{\rm xy}$ as a function of $B_{\perp}$ at different values of $V_{\rm TG}$. Linearity of the transverse resistance indicates single-channel electron transport. (f) Electron Hall mobility $\mu_{e}$ as a function of electron density $n$. Here, $n= B_\perp/e\rho_{\rm xy}$ and $\mu_e=1/(e\rho_{\rm xx} n)$. (g) Electron mean free path $l_e$ as a function of $V_{\rm TG}$. Here, $l_e=(\hbar \mu_e/e)\sqrt{2\pi n}$.}
\end{figure*}

\end{document}